\documentclass[journal]{IEEEtran}
 \pdfoutput=1

\usepackage{cite}
\usepackage{graphicx}
\usepackage{multirow}
\usepackage{amsmath}
\usepackage{tikz}
\usepackage{mathdots}
\usepackage{yhmath}
\usepackage{cancel}
\usepackage{color}
\usepackage{siunitx}
\usepackage{ragged2e}
\usepackage{array}
\usepackage{multirow}
\usepackage{amssymb}
\usepackage{tabularx}
\usepackage{booktabs}
\usetikzlibrary{fadings}
\usetikzlibrary{patterns}
\usetikzlibrary{shadows.blur}
\ifCLASSINFOpdf
\else
\fi

\begin{document}

\title{Evolution of MAC Protocols in the Machine Learning Decade: A Comprehensive Survey}

\author{
Mostafa Hussien, Islam A.T.F. Taj-Eddin, Mohammed F. A. Ahmed, \\ Ali Ranjha, Kim Khoa Nguyen, \and Mohamed Cheriet

%\thanks{This work was funded by Ciena and MITACS Canada with grant number IT13947. M. Hussien, Kim Khoa Nguyen, and Mohamed Cheriet are with École de Technologie Supérieure, Montreal, Canada. H3C 1K3. M. Hussien is with the  Department of Information Technology, Assiut University, Egypt. 71515. (emails: m.korashy@ieee.org)}

\thanks{Manuscript is under processing and not final.}}
\markboth{IEEE Communication Surveys \& Tutorials} {Shell \MakeLowercase{\textit{et al.}}: Bare Demo of IEEEtran.cls for IEEE Journals}

\maketitle

\begin{abstract}
The last decade, (\textbf{2012 - 2022}), saw an unprecedented advance in machine learning (ML) techniques, particularly deep learning (DL). As a result of the proven capabilities of DL, a large amount of work has been presented and studied in almost every field. Since 2012, when the convolution neural networks have been reintroduced in the context of \textit{ImagNet} competition, DL continued to achieve superior performance in many challenging tasks and problems. Wireless communications, in general, and medium access control (MAC) techniques, in particular, were among the fields that were heavily affected by this improvement. MAC protocols play a critical role in defining the performance of wireless communication systems. At the same time, the community lacks a comprehensive survey that collects, analyses, and categorizes the recent work in ML-inspired MAC techniques. In this work, we fill this gap by surveying a long line of work in this era. We solidify the impact of machine learning on wireless MAC protocols. We provide a comprehensive background to the widely adopted MAC techniques, their design issues, and their taxonomy, in connection with the famous application domains. Furthermore, we provide an overview of the ML techniques that have been considered in this context. Finally, we augment our work by proposing some promising future research directions and open research questions that are worth further investigation.
\end{abstract}


\IEEEpeerreviewmaketitle

%#########################################
\section{Introduction}
\label{sec:introduction}

\begin{table}[t]
\begin{tabular}{l|l}
\hline \hline
\multicolumn{1}{c|}{Accronym} & \multicolumn{1}{c}{Meaning}                          \\ \hline \hline
ANN                            & Artificial Neural Networks                          \\ \hline
BSN                            & Body Sensor Network                                 \\ \hline
CR                             & Cognitive Radio                                     \\ \hline
CAP                            & Contention Access Period                            \\ \hline
COP                            & Contention Only Period                              \\ \hline
CSMA/CA                        & Carrier Sensing Multiple Access/Collision Avoidance \\ \hline
CFS                            & Correlation Feature Selection                       \\ \hline
DRL                            & Deep Reinforcement Learning                         \\ \hline
DL                             & Deep Learning                                       \\ \hline
DCF                            & Distributed Coordination Function                   \\ \hline
DSA                            & Dynamic Spectrum Access                             \\ \hline
HDP-HMM                        & Hierarchical Dirichlet Process Hidden Markov Model  \\ \hline
HS                             & Harmony Search                                      \\ \hline
IPI                            & Inter-Packet Interval                               \\ \hline
IDS                            & Intrusion Detection System                          \\ \hline
KNN                            & K Nearest Neighbour                                 \\ \hline
LA-MAC                         & Load Adaptive Medium Access Control                 \\ \hline
LSPI                           & Least Square Policy Iteration                       \\ \hline
MANET                          & Mobile Ad-hoc NETworks                              \\ \hline
MAC                            & Medium Access Control                               \\ \hline
M2M                            & Machine to Machine                                  \\ \hline
MDP                            & Markov Decision Process                             \\ \hline
ML                             & Machine Learning                                    \\ \hline
MAB                            & Multi-Armed Bandit                                  \\ \hline
NN                             & Neural Networks                                     \\ \hline 
NLP                            & Natural Language Processing                         \\ \hline
PDR                            & Packet Delivery Ratio                               \\ \hline
PU                             & Primary User                                        \\ \hline
PDS                            & Post Decision State                                 \\ \hline
QoS                            & Quality of Service                                  \\ \hline
QoE                            & Quality of Experience                               \\ \hline
QNN                            & Q-Neural Network                                    \\ \hline
RTT                            & Round Trip Time                                     \\ \hline
RL                             & Reinforcement Learning                              \\ \hline
RSS                            & Received Signal Strength                            \\ \hline
RF                             & Random Forest
         \\ \hline
SVM                            & Support Vector Machines                             \\ \hline
SMO                            & Sequential Minimal Optimization                     \\ \hline
SAML                           & Self Adapting MAC Layer                             \\ \hline
SU                             & Secondary User                                      \\ \hline
TOP                            & Transmission Only Period                            \\ \hline
TSCH                           & Time-Slotted Channel Hopping                        \\ \hline
TD                             & Temporal Difference                                 \\ \hline
TS                             & Thompson Sampling                                   \\ \hline
TDMA                           & Time Division Multiple Access                       \\ \hline
WMN                            & Wireless Mesh Networks                              \\ \hline
WSN                            & Wireless Sensors Networks                           \\ \hline
WT                             & Wireless Terminal                                   \\ \hline
\hline
\end{tabular}
\end{table}

\IEEEPARstart{D}{uring} the last decade, Machine Learning (ML) algorithms, especially deep learning (DL), have shown an outstanding performance in almost all fields such as computer vision \cite{hassaballah2020deep}, image processing \cite{jiao2019survey}, natural language processing (NLP) \cite{otter2020survey}, etc \cite{lecun2015deep}. This increased success encouraged the researchers to focus on unleashing the power of ML for various problems and domains. Unsurprisingly, ML was continuously breaking new records in all of these fields. Different success stories have been recorded for ML, from \textit{AlexNet}, a Convolution Neural Network (CNN) model which won the \textit{ImageNet} competition \cite{krizhevsky2017imagenet}, to \textit{AlphaGo}, a reinforcement learning model introduced by \textit{DeepMind} in 2016 \cite{silver2016mastering}, and many other ground-breaking models.

The area of wireless communications has been heavily affected by recent advances in ML algorithms and technologies. Specifically, ML was a main player in the 5G and beyond communication systems. It has been adopted to provide modern communication systems with a plethora of intelligent services and functions such as intelligent resource allocation \cite{luong2021deep, wang2021learning}, link adaptation \cite{bobrov2021massive, hussien2021towards}, beamforming \cite{huang2019fast, hojatian2021unsupervised}, automatic modulation classification \cite{meng2018automatic, wang2020lightamc}, channel estimation \cite{soltani2019deep, hu2020deep}, or CSI compression \cite{hussien2022self, liu2020efficient, hussien2020prvnet}. Recently, the advances in ML/DL techniques have also affected the design of medium access control (MAC) protocols \cite{ye2020deep}. Designing an efficient MAC technique is a complex and challenging task since it counts for many factors such as collision, throughput, fairness, hidden terminal problems, etc. With the accelerated developments in 5G towards the 6G, it becomes essential to develop MAC protocols that support different applications scenarios such as ultra-reliable low-latency communications (URLLC) \cite{ranjha2022urllc, narsani2022leveraging}, enhanced wide broadband \cite{abdullah2021enhanced} and massive machine-type communications \cite{dawy2016toward}. Therefore, ML-inspired techniques have great potential for improving and extending the existing MAC mechanisms. As a result, it became an active area of research that attracted many researchers from both academia and industry.

\begin{figure}[t]
    \centering
    \includegraphics[width=0.4\textwidth]{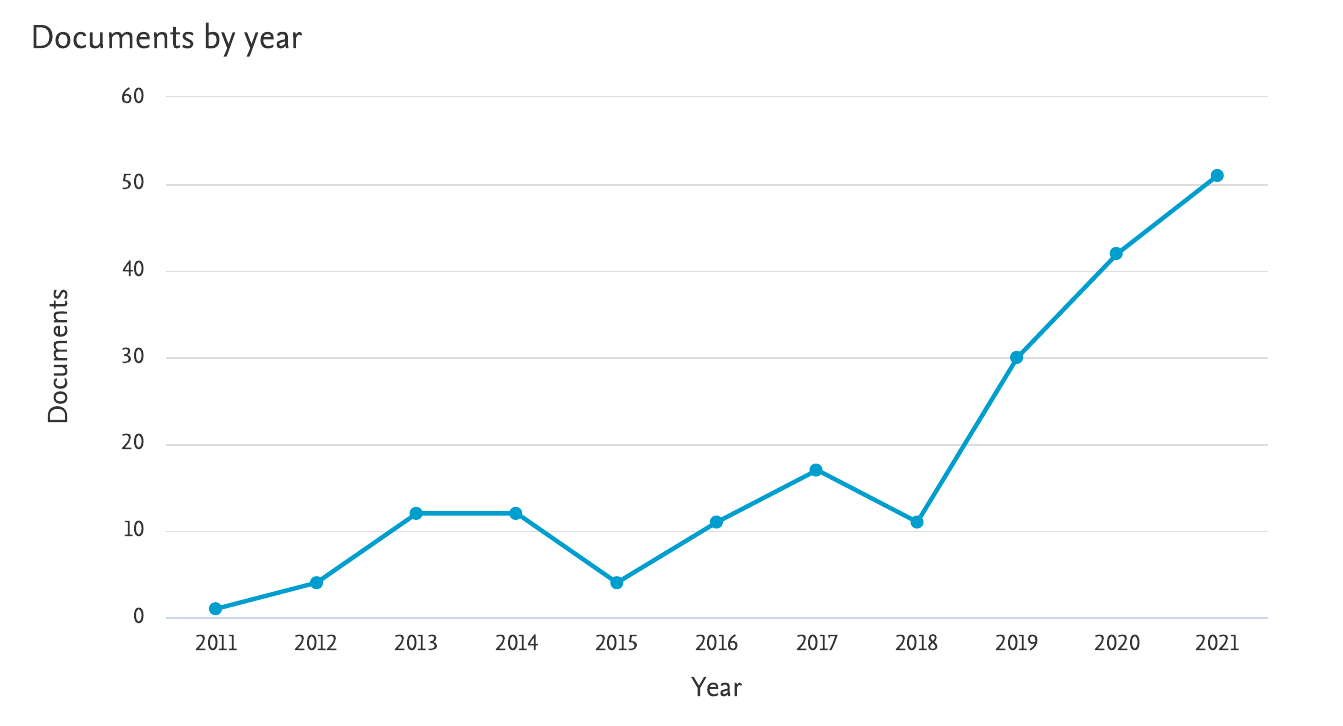}
    \caption{A trend in the number of publications in ML-based MAC protocols, as reported in the \textit{Scopus} database.}
    \label{fig:Scopus_trend}
\end{figure}

The rising interest in ML-based MAC techniques can be clearly observed by the number of articles in this field. The results shown in Fig. \ref{fig:Scopus_trend} were returned when searching the \textit{Scopus} database with the query text \textit{"Machine learning for medium access techniques"}. The figure plots the number of documents per year returned by the query. We can see an exponentially growing trend in the number of articles. Therefore, surveying and classifying the recent work in this area along with a comprehensive review of the widely adopted ML/DL algorithms as well as the MAC design issues can significantly help the researchers accelerate their work.

Different survey papers for MAC protocols have been proposed to summarize the various and latest directions. The survey in \cite{naik2004survey} reviewed and categorized different MAC protocols depending on their medium access strategy (e.g., resource allocation, random access, etc). The authors in \cite{bachir2010mac} reviewed various MAC protocols according to the problem being addressed. The authors in \cite{binti2017optimization} surveyed the work published in the period from 2002 to 2011 in the area of MAC protocols for WSN. Table. \ref{table:other_surveys} shows a comparison between the different surveys and tutorials in this area. 

The following are the main points that distinguish this work from the existing surveys:
\begin{itemize}
    \item We survey the recent work that has been published during a very active period of time (i.e., 2012-2022).
    \item We provide a focus on ML-based MAC techniques only. This makes this survey more specialized for the impact of ML techniques on the problem of medium access. This is an advantage for those who are interested not in the general problem of medium access, but rather in the application of machine learning in this specific problem.
    \item Unlike prior work that focused on classifying the surveyed work from one dimension, we provide a multidimensional classification such that the reader can easily reach the work by their year, the adopted ML technology, or the application domain.
    \item For the sake of completeness, we comprehensively explain the various MAC technique, their design issues, and the relevant ML algorithms. This puts together all the pieces of the game in one place, which facilitates the reading of the survey and the reader does not need to look for other resources that explain the basic concepts and terminologies.
%    \item  We conclude our study by proposing various interesting future research directions and problems. The interested reader can start from one of these points to divide more on the topic of ML-based MAC techniques.
\end{itemize}
Readers can find that survey interesting and can stimulate future research directions and problems on the topic of ML-based MAC techniques.

\begin{table*}[t]
\caption{Comparison between other surveys in the literature and the proposed surveys}
\centering
\label{table:other_surveys}
\begin{tabular}{c|c|c|c|c|c}
\hline \hline
\multicolumn{1}{l|}{} & \multicolumn{1}{l|}{Publication Year} & \multicolumn{1}{l|}{Survey Topic} & \multicolumn{1}{l|}{Employed Technology} & \multicolumn{1}{l|}{Application Domain} & \multicolumn{1}{l}{Time Period Covered} \\ \hline
Huang et al. \cite{huang2012evolution}           & 2012                                  & MAC Protocols                     & General                                  & WSN                                    & 2002-2011                              \\ \hline
Binti et al. \cite{binti2017optimization}          & 2017                                  & MAC Protocols                     & Machine Learning                         & WSN                                     & Unspecified                              \\ \hline
Isolani et al. \cite{isolani2018survey}        & 2018                                  & MAC                               & General                                  & Wireless Networks                       & Unspecified                              \\ \hline
Quintero et al. \cite{quintero2018improvements}       & 2018                                  & MAC                               & General                                  & WSN                                     & Unspecified                              \\ \hline
Zhang et al. \cite{zhang2019deep}           & 2019                                  & General                           & Deep Learning                            & Wireless Networks                       & Unspecified                              \\ \hline
Sun et al. \cite{sun2019application}            & 2019                                  & General                           & Machine Learning                         & Wireless Networks                       & Unspecified                              \\ \hline
Fatima et al. \cite{djiroun2016mac} & & & & & \\ \hline
Ours                   & 2020                                  & MAC                               & Machine Learning                         & Wireless Networks                       & 2011-2021                                \\ \hline \hline
\end{tabular}
\end{table*}

\subsection{Navigation: How you can read this paper?}
\label{subsec:navigation}

\begin{figure}[t]
    \centering
    \includegraphics[width=0.50\textwidth]{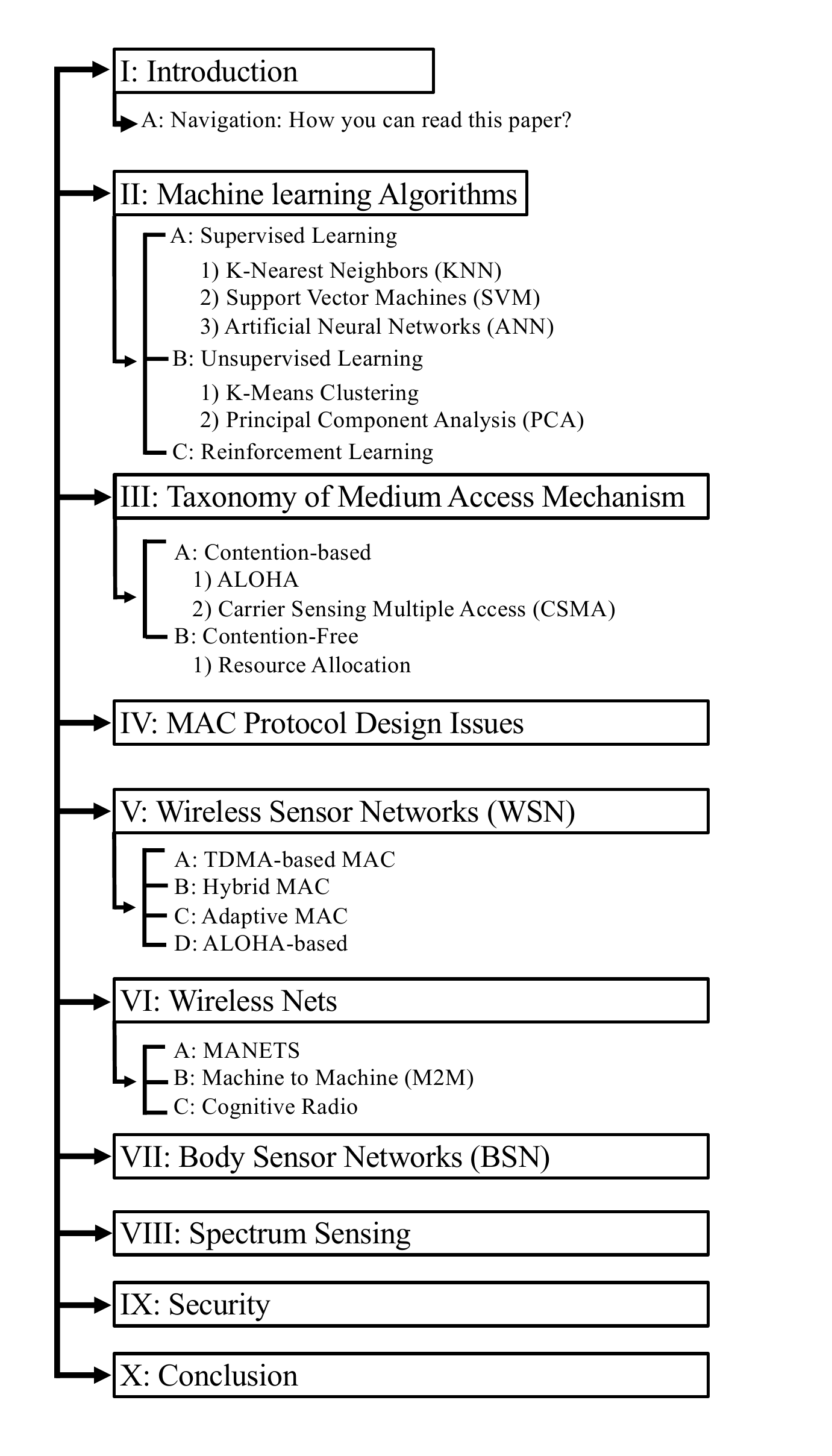}
    \caption{The layout of the paper.}
    \label{fig:paper_layout}
\end{figure}

Navigating through this survey has many forms depending on the objective of the reader. For those who are interested in gaining a piece of knowledge on the underlying concepts and terminologies of ML, they can start with section \ref{sec:machine_learning}. For the readers who are not familiar with the MAC techniques,  \ref{Sec:taxonomy_of_MAC} and \ref{sec:MAC_design_issues} sections give a good background and place the reader in the context of MAC technique design. The surveyed work is organized in sections \ref{sec:wireless_Sensor_nets}, \ref{sec:wireless_nets}, \ref{sec:body_sensor_net}, \ref{sec:spectrum_sensing}, and \ref{sec:security}. Fig. \ref{fig:paper_layout} provides the whole layout of this work. With the help of Fig.\ref{fig:paper_layout}, the reader can easily navigate to the various sections depending on his/her own interest. Moreover, Table \ref{tab:my-table} and Table \ref{tab:Taxonomy} list the surveyed work along with the objective of the work, adopted learning category, and application domain.

%#########################################
%#########################################
%        Machine learning Algorithms
%#########################################
\section{Machine learning Algorithms}
\label{sec:machine_learning}

\begin{figure*}[t]
\centerline{\includegraphics[width=0.9\textwidth]{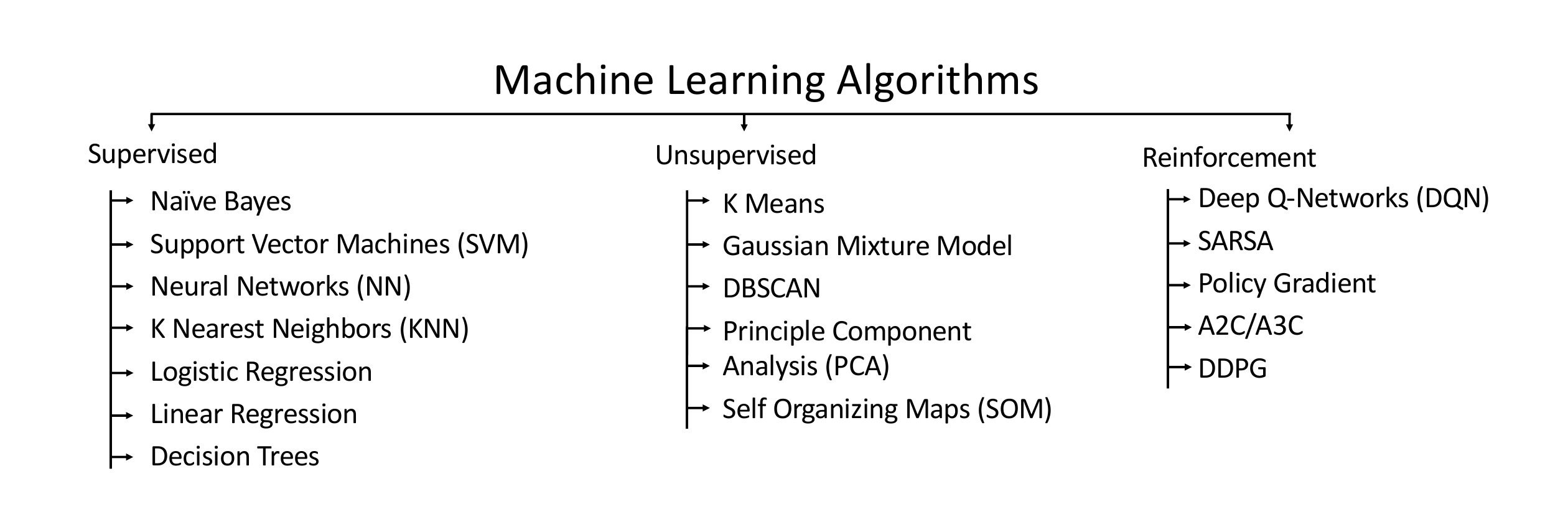}}
\caption{Different algorithms under the umbrella of supervised learning. The figure will include the supervised, unsupervised, and reinforcement learning branches.}
\label{fig:ML_taxonomy}
\end{figure*}

This section covers the various ML algorithms employed in the MAC problem. The employed algorithms belong to one of the three main branches of ML, namely, supervised, unsupervised, and reinforcement learning. We divide this section into three subsections, one for each of the main branches:

\begin{figure}[t]
\centerline{\includegraphics[width=0.45\textwidth]{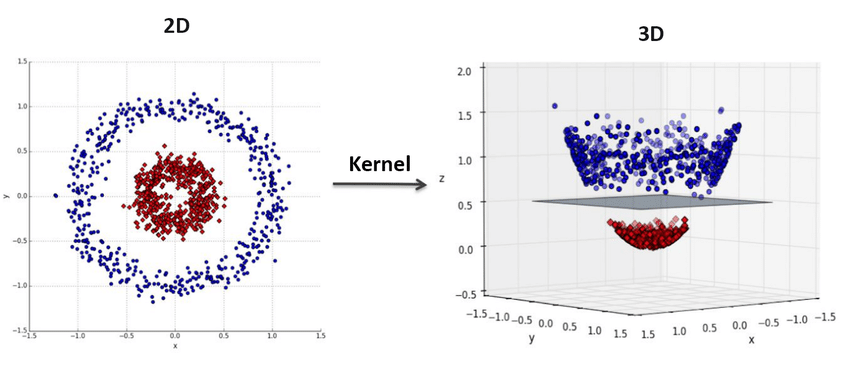}}
\caption{A non-linearly separable data in the input space can be linearly separable in a higher dimensional space.}
\label{fig:kernel_trick}
\end{figure}

%--------------------------------------------
\subsection{Supervised Learning}
\label{subsec:supervised_learning}

Supervised learning is the learning paradigm in which a model learns by examples. This involves showing the model many solved examples of the same problem in the form of (input, output) pairs, i.e. labeled datasets. The model then learns how to map its input to the desired output. Based on the possible values of the output can take, supervised learning can be divided into classification models (discrete output), and regression models (continuous output) \cite{hussien2021fault}. Under each of the classification and regression categories, there are many algorithms as shown in Fig. \ref{fig:ML_taxonomy}.

Among these algorithms, we will cover three of the most adopted algorithms, namely, K-nearest neighbors, support vector machines \cite{yang2010mac,hu2014mac,hu2012mac}, and neural networks:

\subsubsection{K-Nearest Neighbours}
\label{subsubsec:KNN}

K-nearest neighbors (KNN) is one of the simplest and most appealing non-parametric models for both classification and regression. The intuition behind KNN is that close data points usually share the same label \cite{fix1951discriminatory, peterson2009k, garcia2008fast}. The closeness of data points is computed using a similarity measurement function.  

\begin{equation}
\label{eq:ecludian}
    d(x,y) =  \sqrt{  \sum_{i=0}^{n}  (x_i - y_i)^2 }.
\end{equation}

The performance of the KNN algorithm is largely affected by many two factors: the considered number of nearest points, $\textit{K}$, and the used similarity metric. Euclidean distance (\ref{eq:ecludian}) and Cosine similarity (\ref{eq:cosine_sim}) are two widely used functions with KNN.

\begin{equation}
\label{eq:cosine_sim}
    d(x,y) =  \frac{  x^Ty}{||x|| \;  ||y|| }.
\end{equation}

Given the $K$ nearest point, the label of the query point is inferred by an aggregation technique, such as majority voting or distance-weighted voting. In majority voting, the $K$ nearest points assign the query point to the majority class. This implies equally weighting all the $K$ points. In distance-weighted voting, each point participates in the final decision based on its distance from the query point \cite{duda2012pattern}. The main concerns about KNN are that the execution time increases with the dataset size \cite{dudani1976distance}, and the high sensitivity for outliers, see Fig. \ref{fig:KNN_and_SVM}.

\subsubsection{Support Vector Machines (SVM)}
\label{subsubsec:SVM}

\begin{figure*}[t]
\includegraphics[width=0.55\textwidth]{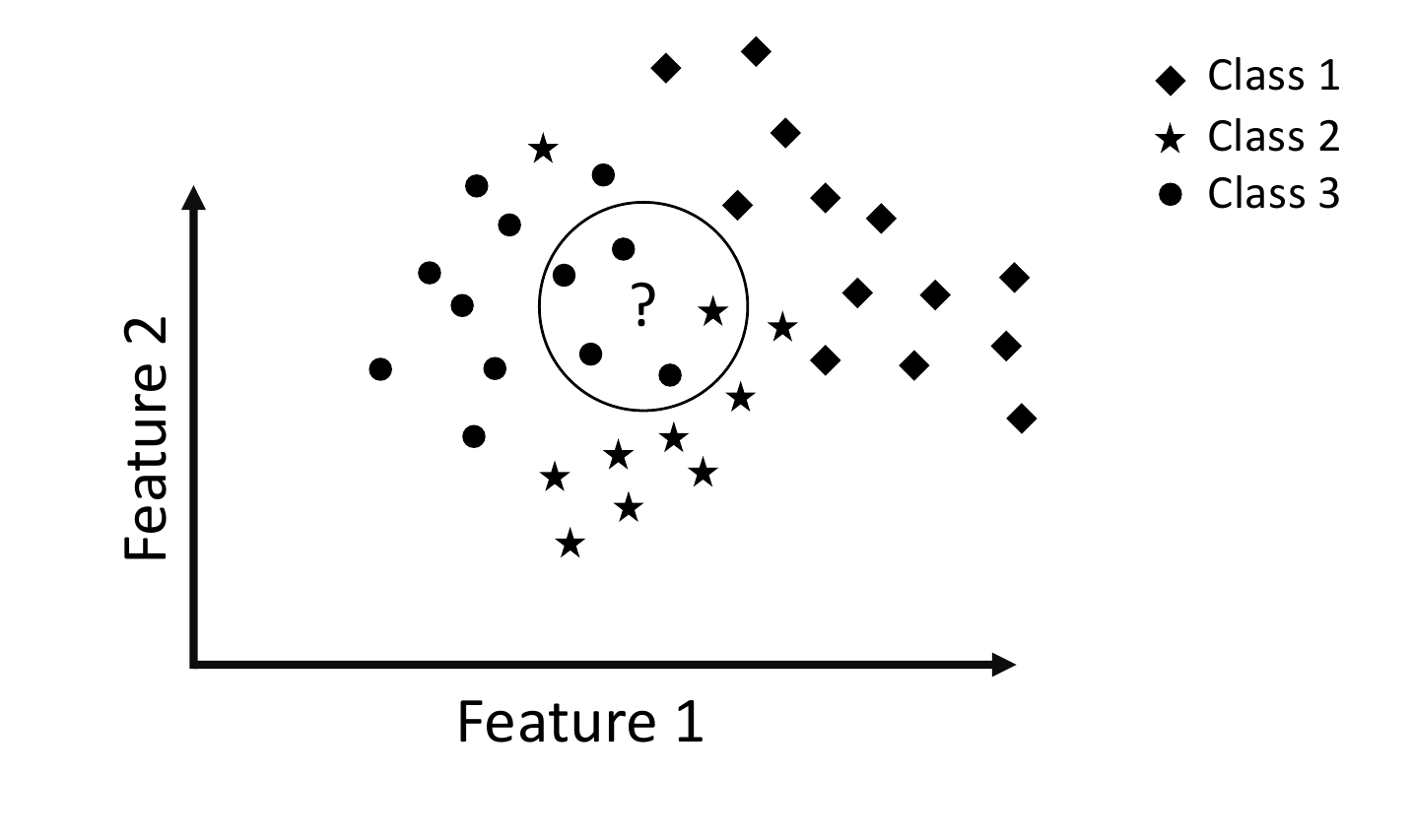}
\includegraphics[width=0.40\textwidth]{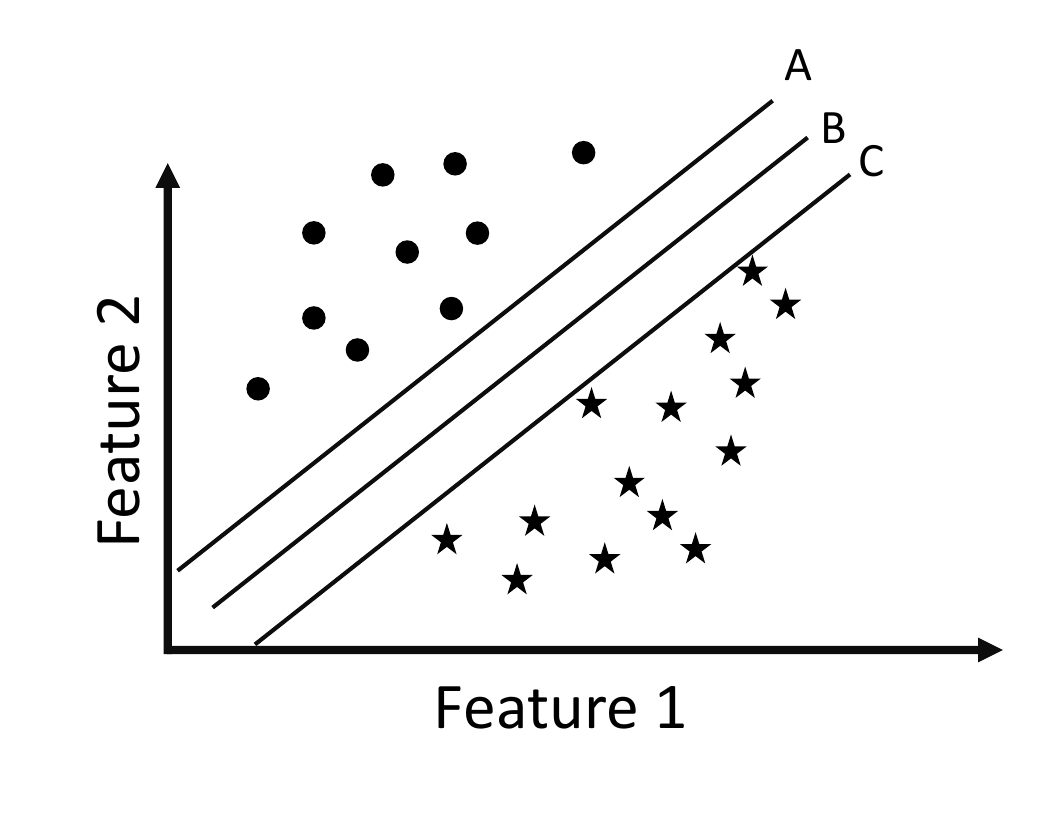}
\caption{An overview of the KNN algorithm for classifying unknown points (left). An overview of the SVM classifier (right).}
\label{fig:KNN_and_SVM}
\end{figure*}

Support Vector Machines (SVM) belong to the family of maximum marginal classifiers. In this family, the model classifies between $N$-dimensional data points by finding an $(N-1)$-hyperplane to split the classes while maximizing the margins between the two classes. In the simplest form, SVM can be seen as a linear classifier. However, in most cases, the data points are not linearly separable. Therefore, they can be mapped to a higher-dimensional space using a kernel function. The idea behind this mapping is that a dataset that is not linearly separable in its original space, maybe linearly separable in higher dimensional space as shown in Fig. \ref{fig:kernel_trick} %re-phrasing 
and Fig. \ref{fig:KNN_and_SVM}. 
For more details on SVM, we refer to \cite{bishop2006pattern}. 

\subsubsection{Artificial Neural Networks}
\label{subsubsec:NN}

Artificial neural networks (ANN), or simply neural networks (NN), is a form of a supervised ML model that can be utilized for both classification and regression problems. The basic building block of an ANN is called a \textit{neuron}, or a \textit{node}. A set of neurons associated with certain inputs is called a layer. The overall neural architecture may consist of one or more layers. Each input to the node is associated with a weight value. The node then computes its output by summing over all weights, the multiplication of each input with its associated weight. An activation function is then applied to the result of the summation as shown in (\ref{eq:neuron_out}).

\begin{equation}
\label{eq:neuron_out}
    f(x) =  \vartheta \left ( \sum_{i=1}^{N} w_i \times x_i \right ),
\end{equation}

\noindent where $N$ is the dimension of the input vector. The activation function, $\vartheta$, takes many forms. \textit{Sigmoid}, $f(x) = 1/1+e^{-x}$, \textit{ReLU}, $\max(0,x)$,  \textit{Tanh}, and $f(x) = \frac{e^{x} - e^{-x}}{e^{x} + e^{-x}}$, are all examples of widely adopted activation functions.

The whole architecture is then trained to minimize a certain loss function. The loss is minimized by updating the weights using an optimization algorithm such as gradient descent, stochastic gradient descent, RMSProp, Adam, etc.

%--------------------------------------------
\subsection{Unsupervised Learning}
\label{subsec:unsupervised_learning}

While in supervised learning, a model learns by example from a labeled dataset, in unsupervised learning, a model learns the underlying structure of an unlabeled dataset. This branch of ML addresses two main problems, namely \textit{clustering} and \textit{dimensionality reduction}. 

As the name implies, clustering algorithms explore the similar underlying structure of a dataset with the aim of grouping similar data points into one group (i.e., cluster). $K$-Means \cite{kanungo2002efficient}, DBSCAN \cite{ester1996density}, and expectation-maximization clustering using Gaussian mixture models (GMM) \cite{moon1996expectation}, are among the most widely used clustering algorithms. 

By contrast, dimensionality reduction is concerned with structuring the data in a way that represents it in a lower-dimensional space as efficiently as possible. Examples of the widely used algorithms for dimensionality reductions are principle component analysis (PCA) \cite{hartigan1979algorithm}, singular value decomposition (SVD) \cite{li2019tutorial}, and linear discriminant analysis (LDA) \cite{izenman2013linear}. 

Following are brief descriptions of the principal component analysis and K-Means clustering algorithms that are often used to address each of the aforementioned problems:

\subsubsection{$K$-Means Clustering}
\label{subsubsec:K_means}

It is an algorithm that aims to partition a number $N$ of data points, into a number $K$ of clusters \cite{hartigan1979algorithm}.  Every data point is assigned to a specific cluster With the nearest mean, cluster center, or cluster's centroid. Based on this interpretation, each cluster's center acts as a prototype for its cluster \cite{kanungo2002efficient}. The algorithm then tries to find the best assignment for the data points that minimizes the intra-cluster differences, given in (\ref{eq:K_means}). 

\begin{equation}
\label{eq:K_means}
    \sum_{j=1}^{K}\sum_{\forall x_i \in c_j}^{}\left \| x_i - \mu_j \right \|.
\end{equation}

In real-world problems, $K$-means has been shown to be effective, but its performance is highly dependent on how the cluster centroids are initialized, as well as the hyperparameter, $K$.

\subsubsection{Principal Component Analysis (PCA)}
\label{SS_Sec:PCA}

\begin{figure}[t]
\centerline{\includegraphics[width=0.45\textwidth]{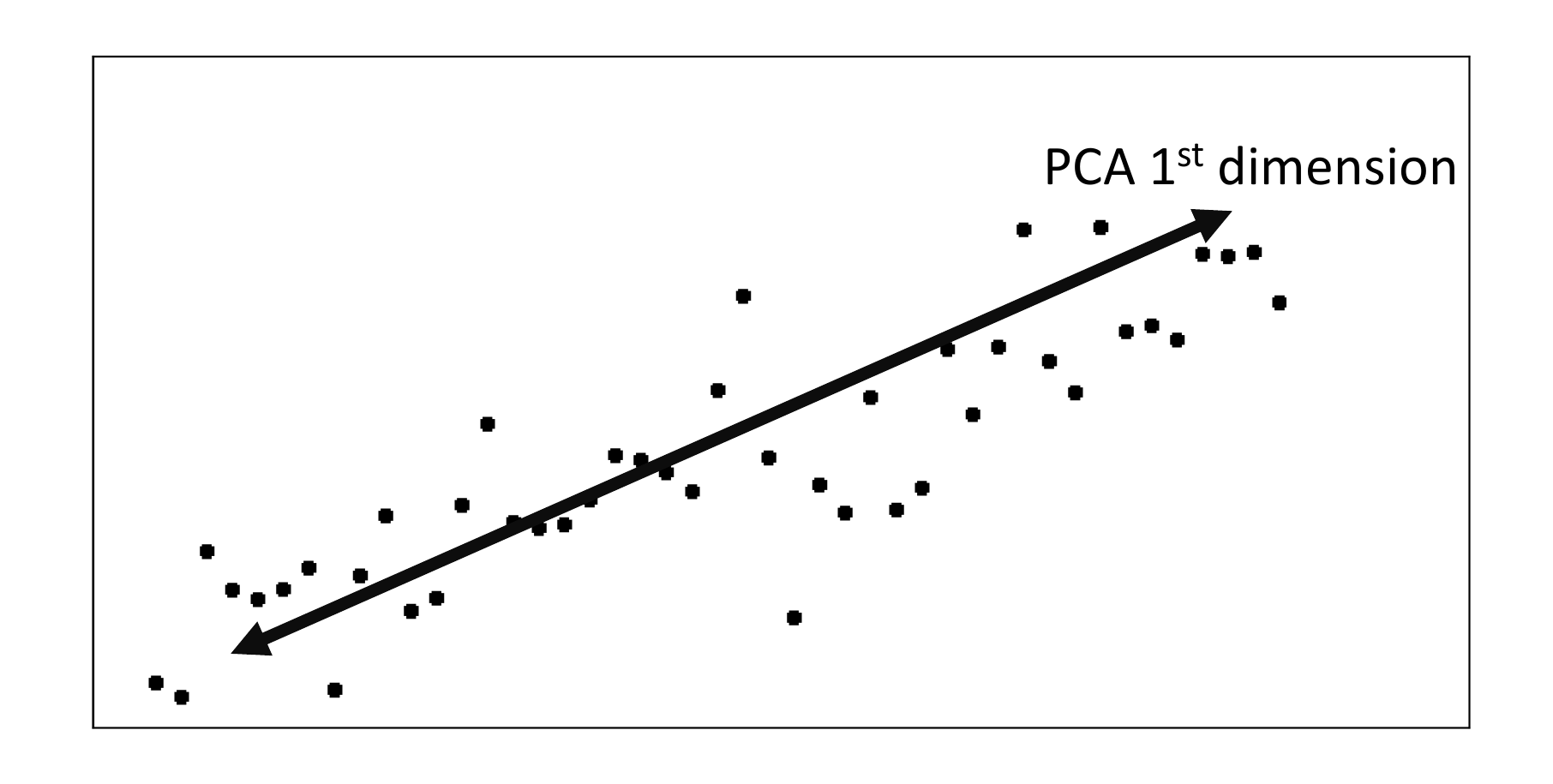}}
\caption{The first principal component is the direction that captures the highest variance between the projected data.}
\label{fig:PCA}
\end{figure}

High-dimensional data is more time-consuming and can affect model learning. Therefore, many ML algorithms and models have problems dealing with high-dimensional data, i.e., the curse of dimensionality \cite{wojtowytsch2020can}. In these cases, we would like to find the best representation of the data in a lower-dimension space. Among the most widely used, and well-studied, approaches for this objective is the principle component analysis (PCA) \cite{ringner2008principal}. PCA can be seen as the process of computing some principal components, some or all of which can be used to change the basis of the data. 

Using the projected data, the first principal component can be seen as the direction capturing the most variance. The second principal component is the direction that captures the second-highest variance for the projected data, etc. A main limitation of PCA is that it can capture linear correlations between the features. If the assumption is not satisfied, PCA can no longer be applied in its current form and further modifications of the algorithm are necessary, see Fig. \ref{fig:PCA}. For more details about PCA and its analysis, we recommend reading \cite{shlens2014tutorial}.

%--------------------------------------------
\subsection{Reinforcement Learning (RL)}
\label{subsec:reinforcement_learning}

Reinforcement learning (RL) is a new paradigm of ML which is different from supervised and unsupervised learning. Supervised and unsupervised ML algorithms find insights in the pre-collected datasets. Specifically, the supervised ML maps the input samples to the corresponding label, while unsupervised ML finds the pattern in this data. On the other hand, RL targets sequential decision-making tasks. RL builds a strategy that helps an intelligent agent to take actions based on the perceived environment. which mimics the human learning process that learns through trial and error \cite{sutton2018reinforcement}.

In RL, there is no learning from a dataset, only a real number or reward signal. The decision-making is done sequentially, and hence time notation is essential in the problem formulation \cite{arulkumaran2017deep}. RL agent, and depending on  interaction with the surrounding environment, takes sequential actions. At each time step, $t$, the agent selects an action, $a_t$, from a set of all possible actions, $A$, which is called the actions space. The environment receives the action, $a_t$, and returns the corresponding reward as well as the next state. The reward signals how well an agent is doing in a given state by letting it know how successful or unsuccessful its actions are. The RL algorithms can be classified into the following classes: \\
1) Value-Based: where the agent tries to maximize a value function, $V(s)$, which is the expected long-term return of a state under a particular policy, as opposed to the short-term reward.\\ 
2) Policy-based: where the agent searches for a rule or policy to take any action in every state to gain maximum reward in the future. The policy is the mapping that the agent learns to determine the next action based on the current state.\\ 
3) Model-Based: where the environment is modeled, and the agent learns to take actions based on that specific model.

%\begin{figure}[t]
%\centerline{\includegraphics[width=0.45\textwidth]{Figs/MAB_RL.pdf}}
%\caption{A description for the difference between Multi-Armed Bandits (MAB), Contextual Multi-Armed Bandits (CMAB), and RL.}
%\label{fig:MAB}
%\end{figure}

%#########################################
\section{Taxonomy of Medium Access Mechanisms}
\label{Sec:taxonomy_of_MAC}

\begin{figure}[t]
    \centering
    \includegraphics[width=\linewidth]{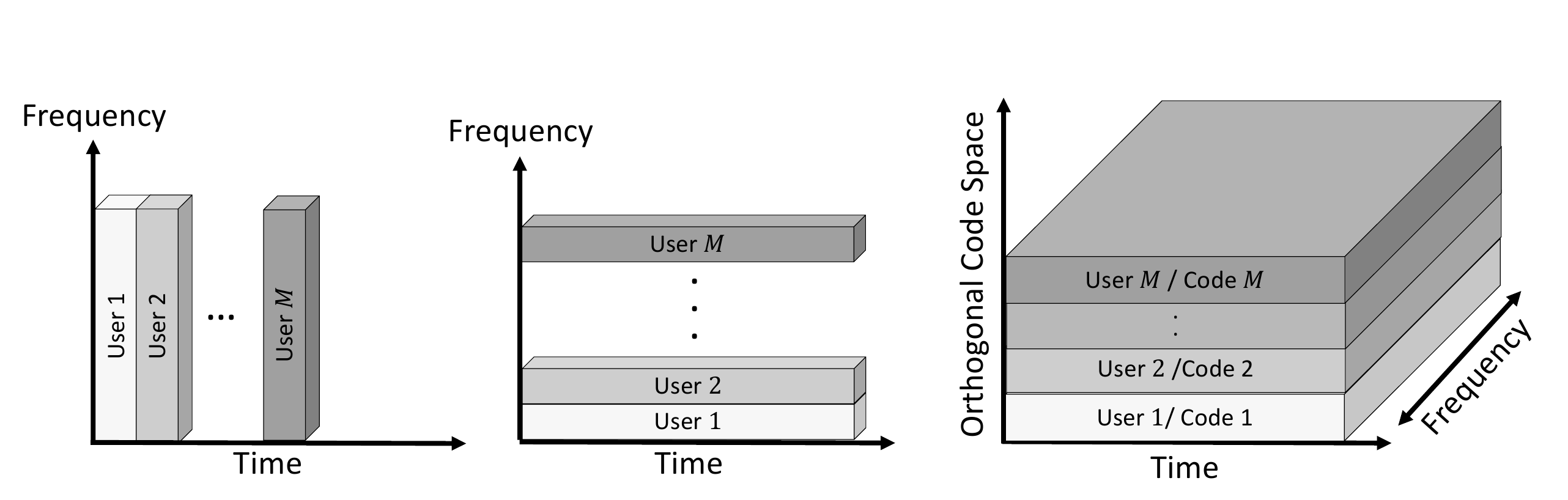}
    \caption{A taxonomy for various medium access mechanisms.}
    \label{fig:MAC_taxonomy}
\end{figure}

Several MAC protocols have been proposed in the literature.
%re-phrasing
According to \cite{djiroun2016mac}, these protocols can be categorized into contention-based and contention-free protocols, see Fig. \ref{fig:MAC_taxonomy}. In this section, we briefly introduce each of these categories and the protocols under each of them.

\subsection{Contention-Based}
A content-based protocol is considered the simplest protocol in terms of setup and implementation. Typically, multiple users can simultaneously communicate on a single radio channel using a contention-based protocol \cite{khisa2020medium}. In this case, several nodes contend to gain access to a shared channel. The main limitation of these protocols is the lack of scalability. Specifically, the number of collisions increases with the number of nodes. The basic contention-based protocols are as follows:

\subsubsection{ALOHA}
%ALOHA protocols attempt to share the channel bandwidth in a more brute-force manner. The original ALOHA protocol was developed as part of the ALOHANET project at the University of Hawaii \cite{abramson1985development}. Strangely enough, the main feature of ALOHA is the lack of channel access control. When a node has a packet to transmit, it is allowed to do so immediately. Collisions are common in such a system, and some form of feedback mechanism, such as an automatic repeat request (ARQ), is needed to ensure packet delivery. When a node discovers that its packet was not delivered successfully, it simply schedules the packet for retransmission. Naturally, the channel utilization of ALOHA is quite poor due to packet vulnerability. The results presented in \cite{roberts1975aloha} demonstrate that the use of a synchronous communication model can dramatically improve protocol performance. This slotted ALOHA forces each node to wait until the beginning of a slot before transmitting its packet. This reduces the period during which a packet is vulnerable to collision and effectively doubles the channel utilization of ALOHA.

%re-phrasing
ALOHA protocols try to share channel bandwidth by using extra brute force manners. The first ALOHA protocol was part of the ALOHAnet project at the University of Hawaii when it was developed \cite{abramson1985development}. Strangely, the lack of channel access control is one of the main features of ALOHA. When a node needs to transmit a packet, it is allowed to be transmitted immediately. In such a system, collisions are usually common.  A feedback mechanism, i.e. automatic repeat request (ARQ), is required to ensure packet delivery. If a node finds that its packet was not successfully delivered, a schedule for the packet to be re-transmitted will be done. Of course, because of packet vulnerability, ALOHA channel utilization is poor. The results presented in \cite{roberts1975aloha} show that to improve protocol performance dramatically, a synchronous communication model can be used. Every node had been forced to wait, by slotted ALOHA, for the start of a slot for transmitting its packet. This way the amount of time a packet is susceptible to collision is reduced, and ALOHA channel utilization doubles effectively.
%-----------------------------------
\subsubsection{Carrier Sensing Multiple Access (CSMA)}

Several MAC protocols used a carrier detection strategy to detect collisions that may happen with ongoing transmissions. First, these protocols listen to the channel to see if there is an ongoing activity. If the channel is free, it will initiate a packet transmission, and if a channel is busy it will suppress it \cite{kleinrock1975packet}.  Persistent CSMA continuously listens, while the channel is busy, to determine when activity stops. The protocol transmits a packet immediately when the channel returns to an idle state. When multiple nodes are waiting for a free channel, Collisions could occur. Randomization had been introduced by Non-persistent CSMA, which will reduce the probability of such collisions. Simply, a source node waits a random backoff time, whenever a busy channel is detected, before retesting the channel. With an exponentially increasing random interval, this process is repeated, until the channel is found clear. The $p$-persistent CSMA protocol represents a mid-point between persistent and non-persistent CSMA. For that case, it is accepted to make the channel slotted, and not to consider the time synchronized. The carrier detection occurs at the beginning of each slot and every slot length is worth the highest propagation delay. A packet with probability $p$ such that $0<p<1$ is transmitted by the node if the channel is free. This procedure will be executed until either the channel becomes busy, or the packet is sent. A source node will be forced to wait a random amount of time, by a busy channel, before starting the procedure again.

Two other well-known variations of CSMA are the Carrier Sensing Multiple Access with Collision Detection (CSMA/CD) and the Carrier Sensing Multiple Access with Collision Avoidance (CSMA/CA). In CSMA/CD, the node continues sensing the channel to detect possible collisions from other nodes' transmissions. CSMA/CA uses a handshaking dialog to minimize the number of exposed nodes and reduce hidden node interference. The mentioned handshake is a control packet of \textit{Request-to-Send (RTS)}  sent to its destination from a source node. A control packet \textit{clear-to-send (CTS)} is the response of the destination, which completes the required handshake. The response from CTS permits the source node to transmit the packet it has. A node is forced to reschedule for later transmission of the packet because of the lack of a CTS \cite{takata2007mac}.

%--------------------------------
\subsection{Contention-Free}
In contention-free protocols, each node does not have to contend to transmit its data. 
%This category can be further divided into two main categories:\\ 
%1) Resource allocation,\\ 
%2) Taking rounds.\\
In the following subsection, we elaborate on the contention-free resource allocation category:
%\begin{figure}[t]
%    \centering
%    \includegraphics[width=\linewidth]%{Figs/MAC_Types.pdf}
%    \caption{Different types of resource allocation-based techniques.}
%    \label{fig:MAC_types}
%\end{figure}

\subsubsection{Resource Allocation}
\begin{figure}[t]
    \centering
    \includegraphics[width=\linewidth]{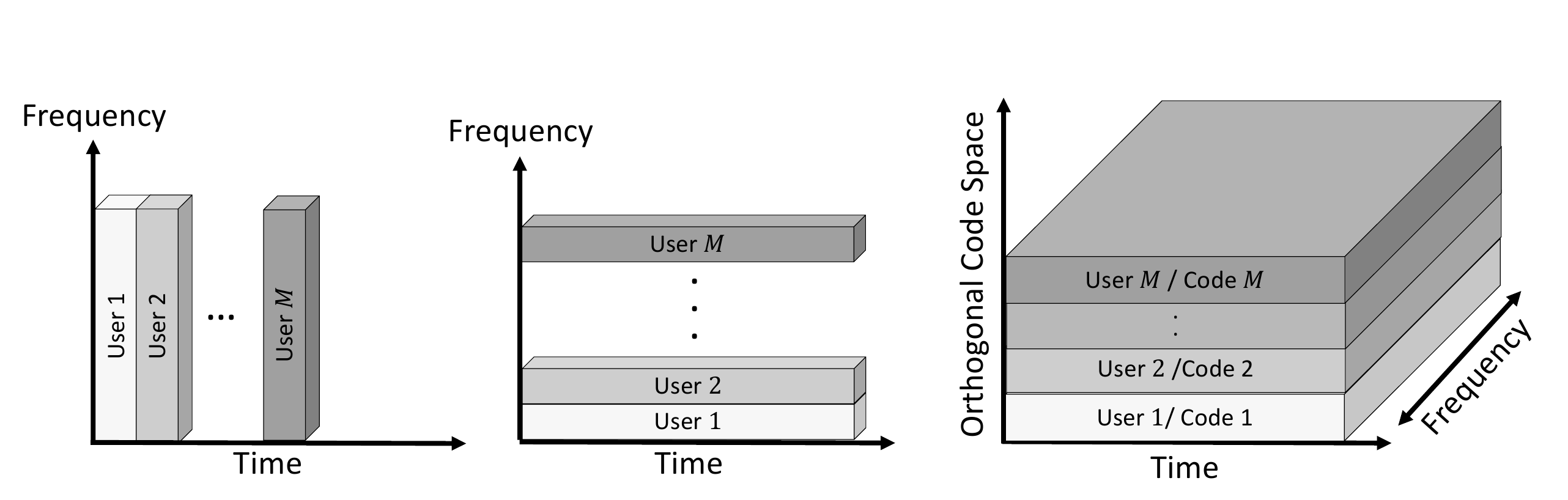}
    \caption{Different types of resource allocation-based techniques.}
    \label{fig:MAC_types}
\end{figure}

In resource allocation-based protocols, each node is assigned a virtual subchannel to transmit its data through, see Fig. \ref{fig:MAC_types}. These virtual subchannels can be time, frequency, or code. Based on the type of the assigned subchannel, we can divide this class into three sub-classes:

\paragraph{Time Division Multiple Access (TDMA)}
%TDMA divides the entire channel bandwidth into $M$ equal time slots that are then organized into a synchronous frame. Conceptually, each slot represents one channel that has a capacity equal to $C/M$ bps, where $C$ is the capacity of the entire channel bandwidth. Each node can then be assigned one (or more) time slot for its exclusive use. Consequently, packet transmission in a TDMA system occurs in a serial fashion, with each node taking turns accessing the channel. Since each node has access to the entire channel bandwidth in each time slot, the time needed to transmit an  $L$-bit packet is then $L/C$. When we consider the case where each node is assigned only one slot per frame, however, there is a delay of $M-1$ slots between successive packets from the same node.

%re-phrasing
TDMA divides the total bandwidth of the channel into $M$ slots of equal time.  They are organized into a synchronous frame. Conceptually, every slot represents a channel its capacity equals $C/M$ bps.  The total channel bandwidth capacity is $C$ bps. One, or more, time slots can be allocated for every node to be exclusively used. Thus, for a TDMA system, packet transmission happens in a serial fashion, and the accessing of the channel happens in turns for every node. The time of $L/C$ is required for transmitting the $L$-bit packet since every node has access to the whole channel bandwidth in every time slot. However, for the same node, a delay of $M-1$ slots between consecutive packets will happen in case every node is allocated one slot per frame exactly.

\paragraph{Frequency Division Multiple Access (FDMA)}
%FDMA divides the entire channel bandwidth into equal subchannels that are sufficiently separated (via guard bands) to prevent co-channel interference. Ignoring the small amount of frequency lost to the guard bands, the capacity of each subchannel is $C/M$, where $C$ is the capacity associated with the entire channel bandwidth. Each source node can then be assigned one (or more) of these subchannels for its own exclusive use. The main advantage of FDMA is the ability to accommodate simultaneous packet transmissions (one on each subchannel) without collision. However, this comes at the price of an increased packet transmission time that results in a longer packet delay.

%re-phrasing

To keep interference from happening to the co-channel, FDMA divides the total bandwidth of the channel into equal sub-channels, that are separated through guard bands. $C/M$ is every sub-channel capacity and $C$ is the total channel bandwidth allocated capacity, assuming that the little amount of frequency that had been lost to the guard bands is neglected. One, or more, of these sub-channels, could be assigned for each source node for its own exclusive use.  The main advantage of FDMA is the capacity to allow concurrent packet transfers, one on every sub-channel, with no collision. This increased flexibility comes with the cost of increased packet transmission time, which results in a bigger packet delay.

\paragraph{Code Division Multiple Access (CDMA)}
%While FDMA and TDMA isolate transmissions into distinct frequencies or time instants, CDMA allows transmissions to occupy the channel at the same time without interference. Collisions are avoided through the use of special coding techniques that allow the information to be retrieved from the combined signal. As long as two nodes have sufficiently different (orthogonal) codes, their transmissions will not interfere with one another. CDMA works by effectively spreading the information bits across an artificially broadened channel. This increases the frequency diversity of each transmission, making it less susceptible to fading and reducing the level of interference that might be caused to other systems operating in the same spectrum. It also simplifies system design and deployment since all nodes share a common frequency band. However, CDMA systems require more sophisticated and costly hardware and are typically more difficult to manage.

%re-phrasing
CDMA allows transmissions to occupy the channel simultaneously without interference, while FDMA and TDMA isolate transmissions into different frequencies or times.  Using special coding methods, that make it possible to retrieve the information out of combined signal, collisions are avoided.  If two nodes have enough orthogonal codes, i.e. different codes, their interference with each other during transmissions will not happen. CDMA functions by successfully spreading the bits of information over an artificially widened channel \cite{ranjha2021urllc}. That way, the diversity of frequency for every transmission is increased, which makes it susceptible less to fading and making the interference level reduced which could happen to other systems that operate in a similar spectrum. Thereby, the design of the system and its deployment are simplified as every node shares a similar frequency band. However, highly sophisticated, expensive, and typically more difficult-to-manage hardware are required for CDMA systems.

\section{MAC Protocol Design Issues}
\label{sec:MAC_design_issues}

% a figure is still needed for this section to illustrate the hidden node problem. you should modify the figure names and place the new figure. 

%\begin{figure}[h]
%    \centering
%    \includegraphics[width=\linewidth]{Figs/HNP.pdf}
%    \caption{The hidden node problem (a) and the exposed terminal problem (b).}
%    \label{fig:hideen_node_prob}
%\end{figure}

%Many protocols have already been proposed for solving the medium access problem in wireless networks \cite{yang2010mac}. When designing a MAC protocol, there are important issues that should be considered. These issues vary from one application to another. For example, in VANETs, the nodes have high movement speed, which concludes in fast and frequent topology changes. While in WiFi networks, the nodes have less mobility and significantly lower speed, which concludes in slower and less frequent topology changes. In this section, we list the most important issues to be taken into account when designing a MAC protocol.

%re-phrasing
Many protocols have already been proposed for solving the medium access problem in wireless networks \cite{yang2010mac}. When designing a MAC protocol, there are important issues that should be considered. These issues vary from one application to another. For example, in VANETs, the nodes have high-speed mobility, resulting in frequent and fast changes in the topology. While in WiFi networks, the nodes have less mobility and significantly lower speed, which concludes in slower and less frequent topology changes. In this section, we list some of the important issues that should be considered when designing a MAC protocol:

\begin{enumerate}
    %\item \textbf{Bandwidth Efficiency:} Due to the limited resource of the radio spectrum, there is limited bandwidth. This bandwidth should be used in an efficient way by decreasing the control overhead, knowing that the bandwidth efficiency can be defined as the ratio of the bandwidth used for data transmission to the total available bandwidth.

    %re-phrasing
    \item \textbf{Bandwidth Efficiency:} Due to the limited resource of the radio spectrum, the bandwidth is limited. To use this bandwidth efficiently, different techniques for decreasing the control overhead should be adopted. Bandwidth efficiency could be defined as the bandwidth ratio used for the transmission of the data to the overall bandwidth available. Increasing this number means we use more bandwidth ratio for data transmission, which is favorable for efficient bandwidth utilization.  
    
    %\item \textbf{Quality-of-Service Support:} Many applications, such as video and voice communication, need some quality of service (QoS) to be guaranteed by the network in order to ensure proper functionality. One good way to provide a good QoS is bandwidth reservation. However, this solution may be difficult to manage in \textit{MANETS} because nodes in such networks are almost mobile.

    %re-phrasing
    \item \textbf{Quality-of-Service Support:} Voice and video communications, among many other applications, require a certain quality of service (QoS) that must be guaranteed by the network operator to ensure proper functionality. Reservation of bandwidth is one way to maintain good QoS. However, since nodes in some networks could be mobile, this solution can be difficult to handle in \textit{MANETS}, for example.
    
    %\item \textbf{Synchronization:} Time synchronization between nodes in a wireless network is very important. In a centralized network, providing a synchronization time is easy to manage since the network has a common clock, the clock of the centralized infrastructure. On the other hand, this is not as easy in \textit{MANETS}, which are totally distributed.
    
    %re-phrasing
    \item \textbf{Synchronization:} In wireless networks, it is of crucial importance to synchronize the operation of different nodes. Reaching this sort of synchronization in centralized networks is easier due to the existence of a centralized clock, i.e., a centralized clock shared by the infrastructure. However, the synchronization is not so straightforward in cases where the control is completely distributed.
    
    \item \textbf{Hidden and Exposed Terminal Problem: }%The hidden terminal problem is easy to understand but not easy to resolve. Fig. \ref{fig:hideen_node_prob} (a) shows three wireless nodes: A, B, and C. A and C cannot communicate directly via their physical layer because they are not within the communication range of each other. But, both of them can communicate with B, which is in their communication range at the same time. Now, suppose A is transmitting data to B. At that time, C cannot hear this transmission; thus, it can transmit at any time, which can cause a transmission collision on B with the ongoing transmission from A. This is what we call the hidden terminal problem, where A and C are each hidden from the other.
    %re-phrasing
    It is easy to understand the problem of the hidden terminal but not easy to resolve it. Three wireless nodes: A, B, and C are shown in Fig. \ref{fig:hideen_node_prob} (a). Since A and C are not within communication range of each other, they cannot communicate directly through their physical layer. But both can simultaneously communicate with B, and B exists in their range of communication. Suppose B receives data from A. Simultaneously at this point, this transmission could not be heard by C. The transmission by C can happen at any time, which can result in node B having a transmission collision during the transmission that is ongoing with node A. This is the problem of the hidden terminal, where A and C together are not heard from each other.
    %The exposed terminal problem is similar to the hidden terminal problem in the sense that the problem is caused by the limitation of the communication coverage range and the common medium. Fig. \ref{fig:hideen_node_prob} (b), which explains this problem, shows four wireless nodes: A, B, C, and D. When B is transmitting data to A, C is prevented from transmitting to D, as it believes that it will interfere with the ongoing transmission from B to A. In reality, C can transmit to D without any risk of interfering with the ongoing transmission between B and A. C is the exposed node.
    %re-phrasing
    The exposed terminal problem is like the problem of the hidden terminal.  The cause of the problem is the coverage limitation of communication within the area and shared medium. In Fig. \ref{fig:hideen_node_prob} (b) the problem is explained and shows four wireless nodes: A, B, C, and D. When B sends data to A, C is forced not to send to D because it has confidence that it will disrupt the data transmission happening from node B to node A. But D can receive from C with no risk of disrupting data transmission happening between A and B. C is the exposed node.

    \item \textbf{Error-Prone Shared Broadcast Channel: }%Because of the transmission radio nature, when a source node is receiving a transmission from a sender node, no other node in its neighborhood should transmit; otherwise, a collision can occur. This is because when a node is transmitting, all nodes in its neighborhood hear this transmission. Since it often happens that nodes in the same neighborhood attempt to access the medium at the same time, the transmission collision probability is quite high in wireless networks. So, the role of a MAC protocol is to reduce these communication collisions as much as possible.
    %re-phrasing
    When a sender node transmits to a source node, no other node in the source node neighborhood should be transmitting due to the broadcast radio nature; otherwise, a collision may occur. Because all nodes that exist close to a transmitted node can hear the transmission of that node.  It is often happening that the medium is accessed at the same time by nodes in the same neighborhood, therefore the probability of transmission collisions in wireless networks is quite high. The role of a MAC protocol is to reduce these communication collisions as much as possible.

    \item \textbf{Distributed Nature and No Central Coordination:} %One of the main characteristics of ad-hoc networks is the lack of any centralized infrastructure or any centralized coordination. In such kinds of networks, nodes should interact in a purely distributed way. Thus, the MAC protocol should carry out good control of the channel access, based on some control packet exchanges that may decrease the bandwidth.
    %re-phrasing
    An ad hoc network's key characteristic is the insufficiency of centralized coordination or centralized infrastructure.  Nodes should communicate in an absolutely distributed manner, in such ad hoc networks.  The MAC protocol should perform efficient channel access control depending upon a number of control packet exchanges that may reduce bandwidth.

    \item \textbf{Mobility of Nodes: } %If nodes were stationary, scheduling the channel access would be much easier than with mobile nodes. A MAC protocol should consider the nature of mobility and the speed of nodes to provide the most efficient mechanism.
    %re-phrasing
    Assuming stationary nodes happen, scheduling the channel access would be much easier than with mobile nodes. A MAC protocol should consider the nature of mobility and the speed of nodes to provide the most efficient mechanism \cite{ranjha2019quasi}.

\begin{figure}[h]
    \centering
    \includegraphics[width=0.9\linewidth]{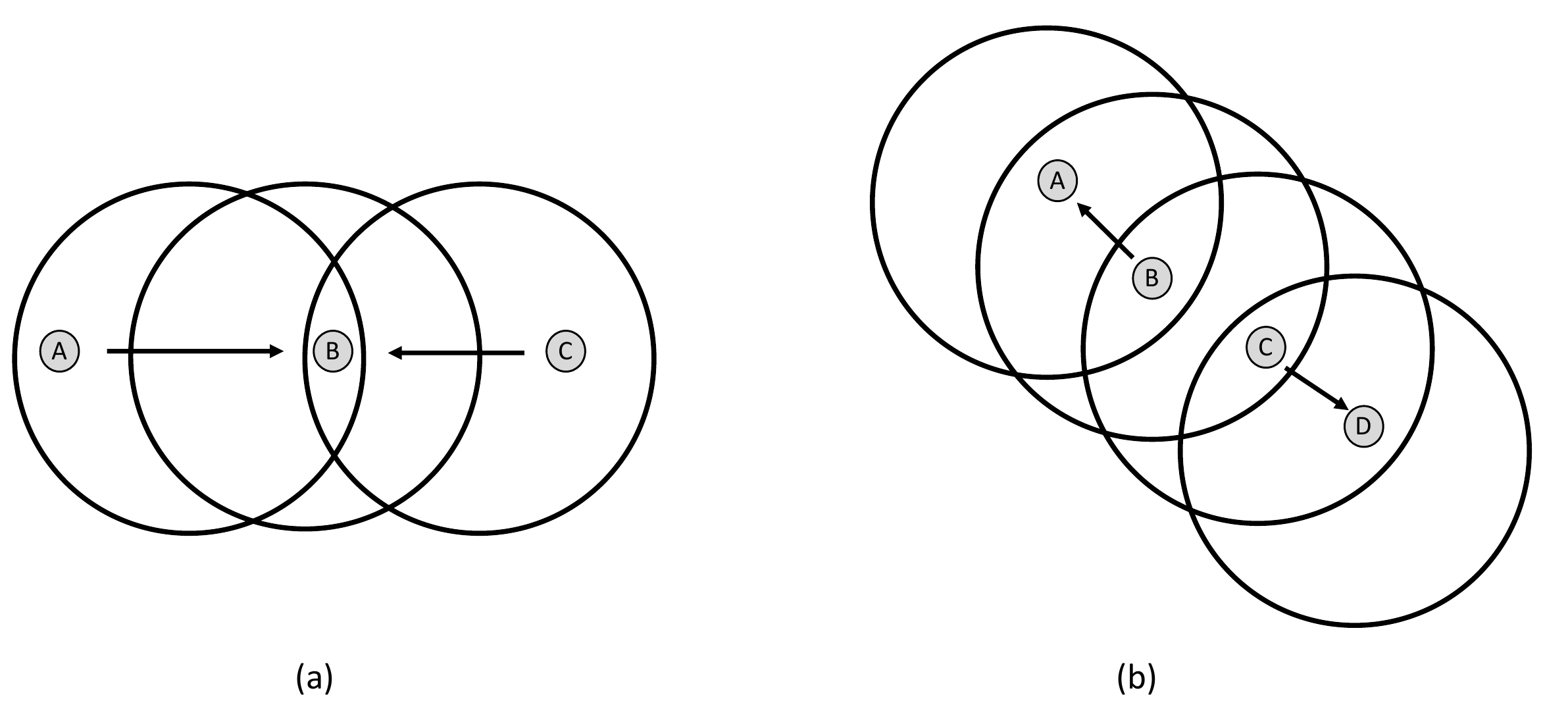}
    \caption{The hidden node problem (a) and the exposed terminal problem (b).}
    \label{fig:hideen_node_prob}
\end{figure}

\end{enumerate}

% \usepackage{graphicx}
\begin{table*}[!t]
\caption{}
\label{tab:my-table}
\resizebox{\textwidth}{!}{
\begin{tabular}{||l||l|l|l|l|l|l|l|l|l||l|l|l|l|l||l|l|l|l|l|l|l|l|l|l|l||}
\hline \hline
   & \multicolumn{9}{l|}{Objective}                                                                                                                                                                                                                                                                                             & \multicolumn{5}{l|}{Employed Technique}                                                                                                                             & \multicolumn{11}{l|}{Application Domain}                                                                                                                                                                                                                                                   \\ \hline \hline
   & \rotatebox{90}{Optimize Power Consumption} & \rotatebox{90}{Maximizing Throughput} & \rotatebox{90}{Maximizing Channel/Spectrum utility} & \rotatebox{90}{reduce control overhead} & \rotatebox{90}{Security (Wireless Intrusion Detection)} & \rotatebox{90}{Supporting real time communications} & \rotatebox{90}{Maximize Video streaming} & \rotatebox{90}{Speed up the learning process for Q-learning algorithms} & \rotatebox{90}{Reduce implementation cost} & \rotatebox{90}{Frame-based procedure} & \rotatebox{90}{MAC Selection for QoS requirements/dynamic circumstances} & \rotatebox{90}{Reinforcement Learning} & \rotatebox{90}{Deep Reinforcement Learning} & \rotatebox{90}{Machine Learning based MAC} & \rotatebox{90}{Heterogeneous IoT Networks} & \rotatebox{90}{Body Sensor Networks} & \rotatebox{90}{M2M Networks} & \rotatebox{90}{Cognitive Radio} & \rotatebox{90}{WSN}                & \rotatebox{90} {Mobile wireless applications} & \rotatebox{90}{Real-Time Wireless Networks} & \rotatebox{90}{Mobile Ad-hoc networks} & \rotatebox{90}{Wireless mesh networks} & \rotatebox{90}{Wireless network virtualization} & \rotatebox{90}{Cognitive Radio ad-hoc networks} \\ \hline
Wang et al. \cite{wang2010cross}  & \checkmark                                &                     &                                     &                         &                                          &                                     &                          &                                                         &                            &                       &                                                          &                        &                             &                           &                            &                      &              &                 &                    &                              &                             &                        &                        &                                 &                                 \\ \hline
Hu et al. \cite{hu2011load}  &                                   & \checkmark                  &                                     &                         &                                          &                                     &                          &                                                         &                            &                       & \checkmark                                                       &                        &                             &                           &                            &                      &              &                 &                    &                              &                             &                        &                        &                                 &                                 \\ \hline
Shah et al. \cite{shah2011ddh}  &                                   &                     &                                     &                         &                                          &                                     &                          &                                                         &                            &                       & \checkmark                                                       &                        &                             &                           &                            &                      &              & \checkmark              &                    &                              &                             &                        &                        &                                 &                                 \\ \hline
Hu et al. \cite{hu2012mac}  & \checkmark                                &                     &                                     &                         &                                          &                                     &                          &                                                         &                            &                       &                                                          &                        &                             &                           &                            &                      &              & \checkmark              &                    &                              &                             &                        &                        &                                 &                                 \\ \hline
Gilani et al. \cite{gilani2013adaptive}  & \checkmark                                & \checkmark                  &                                     &                         &                                          &                                     &                          &                                                         &                            &                       &                                                          &                        &                             &                           &                            &                      &              &                 & \checkmark                 &                              &                             &                        &                        &                                 &                                 \\ \hline
Shah et al. \cite{sha2013self}  &                                   &                     &                                     &                         &                                          &                                     &                          &                                                         &                            &                       & \checkmark                                                       &                        &                             & \checkmark                         &                            &                      &              &                 & \checkmark                 &                              &                             &                        &                        &                                 &                                 \\ \hline
Liu et al. \cite{liu2013scalable}  &                                   & \checkmark                  &                                     &                         &                                          &                                     &                          &                                                         &                            & \checkmark                    &                                                          &                        &                             &                           &                            &                      & \checkmark           &                 &                    &                              &                             &                        &                        &                                 &                                 \\ \hline
Liu et al. \cite{liu2014design}  & \checkmark                                &                     & \checkmark                                  &                         &                                          &                                     &                          &                                                         &                            & \checkmark                    &                                                          &                        &                             &                           &                            &                      & \checkmark           &                 &                    &                              &                             &                        &                        &                                 &                                 \\ \hline
Alvi et al. \cite{alvi2015enhanced}  & \checkmark                                &                     &                                     & \checkmark                      &                                          &                                     &                          &                                                         &                            & \checkmark                    &                                                          &                        &                             &                           &                            &                      &              &                 & \checkmark                 &                              &                             &                        &                        &                                 &                                 \\ \hline
Alvi et al. \cite{alvi2016best} & \checkmark                                &                     & \checkmark                                  & \checkmark                      &                                          &                                     &                          &                                                         &                            & \checkmark                    &                                                          &                        &                             &                           &                            &                      &              &                 & \rotatebox{90}{\checkmark for Smart-Cities} &                              &                             &                        &                        &                                 &                                 \\ \hline
Qiao et al. \cite{qiao2016mac} &                                   &                     &                                     &                         &                                          &                                     &                          &                                                         &                            &                       & \checkmark                                                       &                        &                             & \checkmark                        &                            &                      &              & \checkmark              &                    &                              &                             &                        &                        &                                 &                                 \\ \hline
Ngo et al. \cite{ngo2017user} &                                   & \checkmark                  &                                     &                         &                                          &                                     &                          &                                                         &                            & \checkmark                    &                                                          &                        &                             &                           &                            & \checkmark                   &              &                 &                    &                              &                             &                        &                        &                                 &                                 \\ \hline
Ford et al. \cite{ford2017enhanced} &                                   &                     &                                     &                         &                                          &                                     &                          &                                                         &                            &                       &                                                          &                        &                             & \checkmark                         &                            &                      &              & \checkmark              &                    &                              &                             &                        &                        &                                 &                                 \\ \hline
Yu et al. \cite{yu2019deep} &                                   &                     &                                     &                         &                                          &                                     &                          &                                                         &                            &                       &                                                          &                        & \checkmark                          &                           &                            &                      &              &                 & \checkmark                 &                              &                             &                        &                        &                                 &                                 \\ \hline
Abdulhammed et al. \cite{abdulhammed2018enhancing} &                                   &                     &                                     &                         & \checkmark                                       &                                     &                          &                                                         &                            &                       &                                                          &                        &                             & \checkmark                         &                            &                      &              &                 & \checkmark                 &                              &                             &                        &                        &                                 &                                 \\ \hline
Shoaei et al. \cite{shoaei2019reconfigurable} &                                   &                     & \checkmark                                  &                         &                                          &                                     &                          &                                                         &                            &                       &                                                          & \checkmark                     &                             &                           &                            &                      & \checkmark           &                 &                    &                              &                             &                        &                        &                                 &                                 \\ \hline
Yang et al. \cite{yang2019machine} &                                   &                     &                                     &                         &                                          &                                     &                          &                                                         &                            &                       &                                                          &                        &                             & \checkmark                         & \checkmark                         &                      &              &                 &                    &                              &                             &                        &                        &                                 &                                 \\ \hline
Derakhshani et al. \cite{derakhshani2019self} &                                   &                     &                                     & \checkmark                      &                                          &                                     &                          &                                                         &                            & \checkmark                    & \checkmark                                                       &                        &                             &                           & \checkmark                         &                      &              &                 &                    &                              &                             &                        &                        &                                 &                                 \\ \hline

Chilmulwar et al. \cite{chilmulwar2019novel} & \checkmark                                & \checkmark                  &                                     &                         &                                          &                                     &                          &                                                         &                            &                       &                                                          &                        &                             & \checkmark                         &                            &                      &              &                 &                    & \checkmark                           &                             &                        &                        &                                 &                                 \\ \hline
Franchino et al. \cite{franchino2010energy} & \checkmark                                &                     &                                     &                         &                                          & \checkmark                                  &                          &                                                         &                            &                       &                                                          &                        &                             &                           &                            &                      &              &                 &                    &                              & \checkmark                          &                        &                        &                                 &                                 \\ \hline
Cheng et al. \cite{cheng2010adaptive} &                                   &                     & \checkmark                                  &                         &                                          &                                     &                          &                                                         &                            &                       & \checkmark                                                       &                        &                             &                           &                            &                      &              &                 & \checkmark                 &                              &                             &                        &                        &                                 &                                 \\ \hline
Ju et al. \cite{ju2010scalable} &                                   &                     &                                     & \checkmark                      &                                          &                                     &                          &                                                         &                            &                       &                                                          &                        &                             & \checkmark                        &                            &                      &              &                 &                    &                              &                             & \checkmark                     &                        &                                 &                                 \\ \hline
Naddafzadeh et al. \cite{naddafzadeh2010distributed} &                                   & \checkmark                  &                                     &                         &                                          &                                     &                          &                                                         &                            &                       &                                                          & \checkmark                     &                             &                           &                            &                      &              &                 & \checkmark                 &                              &                             &                        &                        &                                 &                                 \\ \hline
Mihaylov et al. \cite{mihaylov2012decentralised} & \checkmark                                & \checkmark                  &                                     &                         &                                          &                                     &                          &                                                         &                            &                       &                                                          & \checkmark                     &                             &                           &                            &                      &              &                 & \checkmark                 &                              &                             &                        &                        &                                 &                                 \\ \hline
Chu et al. \cite{chu2012aloha} &                                   &                     &                                     & \checkmark                      &                                          &                                     &                          &                                                         &                            &                       &                                                          & \checkmark                     &                             &                           &                            &                      &              &                 & \checkmark                 &                              &                             &                        &                        &                                 &                                 \\ \hline

Changuel et al. \cite{changuel2012online} &                                   &                     &                                     &                         &                                          &                                     & \checkmark                       &                                                         &                            &                       &                                                          & \checkmark                     &                             &                           &                            &                      &              &                 & \checkmark                 &                              &                             &                        &                        &                                 &                                 \\ \hline
Galzarano et al. \cite{galzarano2013ql} & \checkmark                                &                     &                                     &                         &                                          &                                     &                          &                                                         &                            &                       &                                                          & \checkmark                     &                             &                           &                            &                      &              &                 & \checkmark                 &                              &                             &                        &                        &                                 &                                 \\ \hline
Hu et al. \cite{hu2014mac} &                                   &                     & \checkmark                                  &                         &                                          &                                     &                          &                                                         &                            &                       &                                                          &                        &                             & \checkmark                       &                            &                      &              & \checkmark              &                    &                              &                             &                        &                        &                                 &                                 \\ \hline
Rovcanin et al. \cite{rovcanin2014reinforcement} &                                   &                     &                                     &                         &                                          & \checkmark                                  &                          &                                                         &                            &                       &                                                          & \checkmark                     &                             &                           &                            &                      &              &                 & \checkmark                 &                              &                             &                        &                        &                                 &                                 \\ \hline
Al et al. \cite{al2014multi} &                                   & \checkmark                  &                                     &                         &                                          &                                     &                          &                                                         &                            &                       &                                                          & \checkmark                     &                             &                           &                            &                      &              &                 &                    &                              &                             &                        & \checkmark                     &                                 &                                 \\ \hline
Chu et al. \cite{chu2015application} & \checkmark                                & \checkmark                  &                                     &                         &                                          &                                     &                          &                                                         &                            &                       &                                                          & \checkmark                     &                             &                           &                            &                      &              &                 & \checkmark                 &                              &                             &                        &                        &                                 &                                 \\ \hline
Amuru et al. \cite{amuru2015send} &                                   &                     &                                     &                         &                                          &                                     &                          & \checkmark                                                      &                            &                       &                                                          & \checkmark                     &                             &                           &                            &                      &              &                 & \checkmark                 &                              &                             &                        &                        &                                 &                                 \\ \hline
Shoaei et al. \cite{shoaei2015learning} &                                   &                     & \checkmark                                  &                         &                                          &                                     &                          &                                                         & \checkmark                         &                       & \checkmark                                                       &                        &                             &                           &                            &                      &              &                 &                    &                              &                             &                        &                        & \checkmark                              &                                 \\ \hline
Kakalou et al. \cite{kakalou2015reinforcement} &                                   & \checkmark                  &                                     &                         &                                          &                                     &                          &                                                         &                            &                       &                                                          & \checkmark                     &                             &                           &                            &                      &              &                 &                    &                              &                             &                        &                        &                                 & \checkmark                              \\ \hline
Mastronarde et al. \cite{mastronarde2016reinforcement} & \checkmark                                &                     &                                     &                         &                                          &                                     &                          &                                                         &                            &                       &                                                          & \checkmark                     &                             &                           &                            &                      &              &                 &                    & \checkmark                           &                             &                        &                        &                                 &                                 \\ \hline
Qiao et al. \cite{qiao2018intelligent} &                                   &                     & \checkmark                                  &                         &                                          &                                     &                          &                                                         &                            &                       &                                                          &                        &                             & \checkmark                        &                            &                      &              &                 & \checkmark                 &                              &                             &                        &                        &                                 &                                 \\ \hline
Cordeiro \cite{cordeiro2018fs} & \checkmark                                & \checkmark                  & \checkmark                                  &                         &                                          &                                     &                          &                                                         &                            &                       & \checkmark                                                       &                        &                             & \checkmark                         &                            &                      &              &                 & \checkmark                 &                              &                             &                        &                        &                                 &                                 \\ \hline
Xu et al. \cite{xu2020multi} &                                   &                     & \checkmark                                  &                         &                                          &                                     &                          &                                                         &                            &                       &                                                          &                        & \checkmark                          &                           &                            &                      &              &                 & \checkmark                 &                              &                             &                        &                        &                                 &                                 \\ \hline \hline
\end{tabular}
}
\end{table*}

\begin{table*}[t]
  \centering
  \caption{Taxonomy of MAC protocols in the last decade}
  \renewcommand{\arraystretch}{1.5}
  \begin{tabular} {l|c|c|c}
  \hline \hline
    Year & Ref. & Taxonomy & Description \\ \hline 
    2010 & \cite{wang2010cross} & HMAC & Hybrid MAC \\ \hline 
    2011 & \cite{hu2011load} & LA-MAC & Load Adaptive - MAC \\ \hline 
    2011 & \cite{ shah2011ddh} & DDHMAC & Dynamic-Decentralized Hybrid MAC \\ \hline 
    2013 & \cite{ sha2013self } & SAML & Self-Adapting MAC \\ \hline 
    2015 & \cite{ alvi2015enhanced} & BSMAC & Bitmap-Assisted Shortest Job MAC \\ \hline 
    2016 & \cite{alvi2016best} & BEST-MAC & Bitmap-Assisted Efficient \& Scalable TDMA based MAC   \\ \hline 
    2017 & \cite{ngo2017user} & TSCH & Time Slotted Channel Hopping \\ \hline 
    2019 & \cite{yu2019deep} & DLMA & Deep Learning MAC \\ \hline \hline
  \end{tabular}
  \vspace{5pt}
  \label{tab:Taxonomy}
\end{table*}

%####################################################
%       Wireless Sensor Networks (WSN)
%####################################################
\section{Wireless Sensor Networks (WSN)}
\label{sec:wireless_Sensor_nets}
\begin{figure}
    \centering
    \includegraphics[width=8cm]{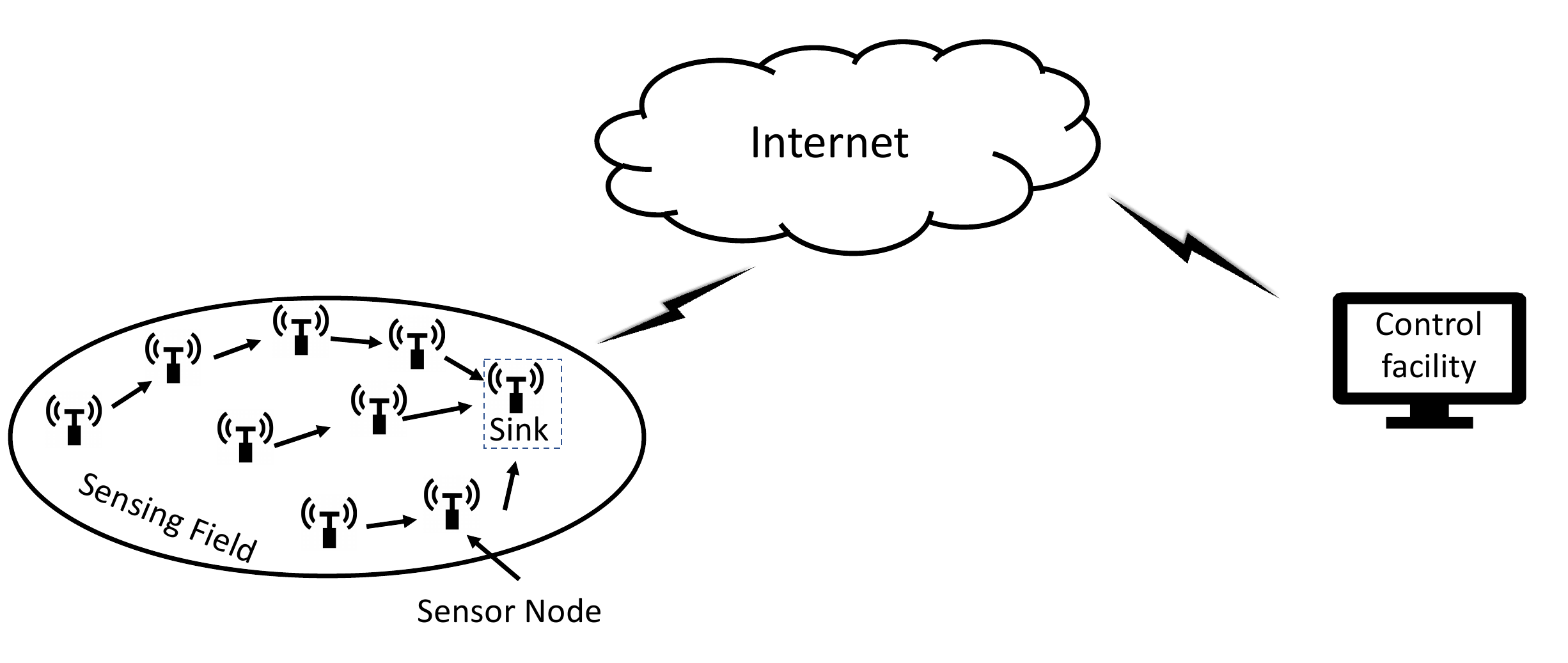}
    \caption{A wireless sensor network (WSN) fusing the sensed data to a control facility through the Internet.}
    \label{fig:WSN_illusteration}
\end{figure}

A WSN is a network that consists of many nodes, these nodes can sense and process data through multi-hop communication. Actually, WSN nodes have different transmission behavior and variations in their traffic loads, see Fig. \ref{fig:WSN_illusteration}. The WSN lifetime mostly depends on the efficiency of energy due to the difficulties of replacing or recharging nodes \cite{ranjha2020quasi}. Other challenges are facing WSNs according to the used MAC protocol, such as increased delay and less throughput \cite{hussien2022learning}. A MAC protocol designed with good quality is required to make sure of successfully delivering the data while keeping the energy consumption at its minimum.

In \cite{mihaylov2012decentralised}, WSNs create a derived collection, i.e. class of networks. Our daily environment could be monitored by this derived collection of networks. Typically, a WSN consists of a set of sensor nodes with limited resources in terms of energy and communication. These limited resources complicate a WSN application design where the application requirements (e.g., throughput, latency, or lifetime) often conflict with the network capacity and the energy resources. To address these challenges, a standard approach is followed. This standard approach depends on scheduling wake-up that alternates between sleep and active states for sensor nodes.
On-demand paging, synchronous, and asynchronous wake-up are the three commonly used wake-up solutions.
\begin{itemize}
    \item With on-demand paging, a separate radio managed the wake-up function, and much less power is used by the separate radio than the master radio when inactive. Therefore, the master radio remains in a state of sleep until a signal came from the slave radio tells that the radio channel is about to receive a message. However, there is an additional effort in hardware design that should not be neglected.
    \item In the synchronous wake-up approaches, the nodes turn on their radios in a coordinated manner, allowing nodes to wake up at predetermined times when communication between nodes becomes possible. The main problem with such a solution is the overhead caused by synchronizing the nodes.
    \item Finally, with asynchronous wake-up solutions, the nodes are unaware of each other's schedules, and the communication involves increased costs for both sender and receiver.    
\end{itemize}

The authors in \cite{mihaylov2012decentralised} 
presented a self-organizing approach using RL to schedule the node's wake-up cycles in a WSN. Two main conditions must be met for nodes to wake up. 
\begin{enumerate}
    \item For communication to succeed, the node's parent in the routing tree needs to be awake. 
    \item The neighboring nodes should not be active except for the parent node. 
\end{enumerate}
Also, the schedules of the neighboring nodes that do not need to communicate are desynchronized to avoid radio interference and packet loss. Nodes in the WSN use an RL algorithm to learn a wake-up schedule, i.e. within the same frame when to remain active, such that in a distributed environment lifetime and throughput will be improved. Each node stores a quality value (Q-value) for each slot within its frame. This value indicates how advantageous for the node to stay awake during these slots for each frame, i.e. given the work cycle of the node and considering node communication history, which is an efficient wake-up pattern. When a communication event occurs at a node, i.e. a packet is listened to, received, or sent, or when at the time of active period events have not occurred, i.e. inactive listening, the Q-value of the slot(s) will be updated by that node.
Experiments are conducted on three network topologies namely line, mesh, and grid topologies with different sizes.  Experimental results show that both synchronous and asynchronous are crucial in WSN applications that demand real-time data. The method proposed in \cite{mihaylov2012decentralised} allowed the coordination of agents in their wake-up cycles with no communication overhead and make sure that neighboring nodes evade interference of communication at no energy cost. To decrease the number of time slots that are active inside a frame and to relax the condition that the nodes' active time in a period that is contiguous, many optimizations are needed.

%---- Table in subsection V.A (TDMA-Based MAC)
% \usepackage{graphicx}
\begin{table*}[t]
\centering
\caption{BS-MAC and BMA-RR based MAC protocol transmission delay of data comparison.}
\label{table:BS_MAC}
\resizebox{\linewidth}{!}{%
\begin{tabular}{|p{0.15in}|p{0.3in}|c|c|c|c|c|c|c|c|c|c|c|c|}
\hline
  \multicolumn{1}{|p{0.15in}|}{Node} &
  \multicolumn{1}{p{0.3in}|}{Data length (bytes)} &
  \multicolumn{1}{p{0.3in}|}{Data rate (bps)} &
  \multicolumn{1}{p{0.42in}|}{Time to send data (ms)} &
  \multicolumn{1}{p{0.4in}|}{Bits/slot in BMA-RR} &
  \multicolumn{1}{p{0.3in}|}{Slot length BMA-RR MAC (ms)} &
  \multicolumn{1}{p{0.2in}|}{Req slots} &
  \multicolumn{1}{p{0.45in}|}{Time to send data in BMA-RR (ms)} &
  \multicolumn{1}{p{0.3in}|}{Bits/slot in BS-MAC} &
  \multicolumn{1}{p{0.3in}|}{Slot length in BS-MAC} &
  \multicolumn{1}{p{0.3in}|}{Slots required} &
  \multicolumn{1}{p{0.45in}|}{Time to send data in BS-MAC (ms)} &
  \multicolumn{1}{p{0.3in}|}{Time lapsed in BMA-RR (ms)} &
  \multicolumn{1}{p{0.3in}|}{Time lapsed in BS-MAC (ms)} \\ \hline
A & 120 & 24,000 & 40    & 2,000 & 83.33 & 1 & 83.33  & 200 & 8.33 & 5  & 41.67 & 43.33 & 1.67 \\ \hline
B & 180 & 24,000 & 60    & 2,000 & 83.33 & 1 & 83.33  & 200 & 8.33 & 8  & 66.67 & 23.33 & 6.67 \\ \hline
C & 210 & 24,000 & 70    & 2,000 & 83.33 & 1 & 83.33  & 200 & 8.33 & 9  & 75.00 & 13.33 & 5.00 \\ \hline
D & 240 & 24,000 & 80    & 2,000 & 83.33 & 1 & 83.33  & 200 & 8.33 & 10 & 83.33 & 3.33  & 3.33 \\ \hline
E & 280 & 24,000 & 93.33 & 2,000 & 83.33 & 2 & 166.67 & 200 & 8.33 & 12 & 100.0 & 73.33 & 6.67 \\ \hline
\end{tabular}%
}
\end{table*}
%---------------------------------------------

In \cite{galzarano2013ql}, a WSN typically consists of small, inexpensive, and low-power devices that provide data acquisition, processing, and wireless communication functions. Network endurance is a characteristic of WSN which is considered very important. Sensor nodes can be deployed in remote areas such that replacing or maintaining batteries is a difficult or impossible task. Access to the shared communication channel is the responsibility of the MAC protocol, also it is intended to ensure that successful radio management happens with efficient energy management. Efficient energy management can be achieved by avoiding the sources that cause waste of energy, i.e. Overhearing, inactive listening, packet collisions, and excessive retransmissions. 
Adaptive behavior that takes into account the actual network conditions is urgently needed to address the network lifetime issue. This research introduces QL-MAC, which is a new contention-based protocol for MAC belonging to WSN.  The proposed protocol employs the Q-learning algorithm to get an effective wake-up strategy to decrease the consumption of energy based on the real load of the network of neighborhoods.  It gains from the interaction of the cross-layer with the network layer to provide an improved understanding of the patterns of communication and hence decrease the consumption of energy because of overhearing and inactive listening. The goal of the protocol proposed is to let the behaviors of nodes be inferred from each other to approve a virtuous active/sleep schedule policy. Typically, RL is a machine learning sub-area specialized in how a user takes actions to increase some reward that will be gained in the long term. Q Learning is the most popular and powerful algorithm based on RL and it does not require the modeling of the environment and its actions depending on a Q function that determines a certain action quality at a certain state of an agent.  In particular, each node has to decide whether it should be in active or in sleep mode during each single time slot. All nodes store a set of Q values, that exist in an explicit slot inside the frame. The Q value provides evidence of the gains a node has when during the relevant time slot, the node is awake. The Q value is modernized over time based on some explicit events that happen through the very slot in every frame.  Simulations are conducted using OMNET++ to compare the performance of the proposed protocol against related MAC protocols in terms of Packet Delivery Ratio (PDR) and nodes' average consumption of energy. Results show that QL MAC outperforms related MAC protocols as the proposed protocol allows nodes to spend much less energy due to the sleep/wake-up radio schedule. Also, the results show that when the slots number decrease, QL MAC generally exhibits higher performance respectfully to the PDR but, the energy spent by the node tends to increase.

A number of MAC protocols have been proposed for WSNs to ensure the successful delivery of data while reducing energy consumption. The proposed protocols have made a great impact on the efficiency of energy and the performance of a channel.  But, with the advances in research, the complexity of MAC protocols is increasing and the overhead of the control packet consumes more resources of a channel. That raised the need for simpler and more efficient MAC protocols. MAC protocols for WSNs might be classified into different categories such as TDMA-based \cite{alvi2015enhanced,alvi2016best},\cite{franchino2010energy}, Hybrid MAC protocols where two different protocols are combined together \cite{wang2010cross,gilani2013adaptive}, adaptive/dynamic MACs
\cite{ sha2013self },
\cite{qiao2018intelligent},
\cite{mastronarde2016reinforcement},
\cite{shoaei2019reconfigurable},
\cite{galzarano2013ql},
%FDMA-based, 
CSMA-based \cite{gilani2013adaptive},\cite{mastronarde2016reinforcement}, and ALOHA-based \cite{chu2012aloha},\cite{chu2015application}.  The coming sub-sections A, B, C, and D, will explain these categories in more detail.
%---------------------------------------------------------
\subsection{TDMA-Based MACs}
\label{subsec:TDMA_MAC}

In \cite{alvi2015enhanced}, the authors present an adaptive TDMA-based MAC protocol, called Bitmap-assisted Shortest job first based MAC (BSMAC) for hierarchical wireless sensor networks. The number of small-sized slots of time that BSMAC takes into account is not the same as the number of member nodes number. The Shortest Job First (SJF) approach is used to decrease the node average delay of a packet and shorten the job time completion of the node. The proposed protocol minimizes the control overhead and energy consumption by using 1 byte short address to identify the member nodes.  
Experiments are conducted to evaluate the proposed Hybrid MAC protocol against existing MAC protocols in terms of throughput, energy efficiency, and delay. The results show that BSMAC outperforms the existing MAC protocols in terms of throughput. Additionally, BSMAC uses less energy than the rest of the protocols of MAC for the equivalent quantity of data. BSMAC has significantly less transmission delay due to the implication of the SJF algorithm as nodes transmit their data at once instead of transmitting in parts. Table \ref{table:BS_MAC} compares the transmission delay of the BS-MAC and BMA-RR based MAC protocol.

In \cite{alvi2016best} the authors present a new TDMA-based MAC protocol, called Bitmap-assisted Efficient and Scalable TDMA-based MAC (BEST-MAC). The primary goal of BEST-MAC is to enhance quality control in applications for smart cities where diversified traffic is in demand and data loss, or delays are undesirable. 
In literature, several TDMA-based MAC protocols have been proposed to overcome certain E-TDMA and TDMA limitations at the expense of higher control overheads. Furthermore, these techniques do not deal with scaling problems as they offer a fixed amount of data slots that are the same number as the participant nodes, which are unable to handle adaptive traffic load efficiently.  
BEST-MAC uses a large number of small-size time slots for a growing Link usage that efficiently adapts to changing traffic demands. The algorithm of Knapsack is employed to shorten the time it takes for a node to complete a task and allows an increase in the number of nodes that may transfer data concurrently. The suggested scheme also adds a distinct conflict access duration to allow non-member nodes to join the network while data is being transferred. Control expenses and energy consumption are both minimized by giving each member node a one-byte short address to pinpoint them.  
Experiments are conducted to evaluate the BEST-MAC protocol against conventional MAC protocols in terms of throughput, energy efficiency, and delay. The results show that BEST-MAC significantly improves throughput because of the use of the optimization approach of knapsack and the choice of smaller slots of data.  Also, results show that BEST-MAC has less transmission delay and use of energy in comparison to the current protocols of MAC. In literature, real-time communication and energy saving over wireless networks have received much attention. However, addressing both problems simultaneously is rarely considered. 

In \cite{franchino2010energy}, the authors proposed an energy model to reduce energy consumption in wireless networks with real-time requirements. In particular, the paper presents El-MAC, an elastic energy-aware algorithm to save energy at the communication level, precisely at the MAC level where each node can adapt its bandwidth requirements to balance performance versus energy consumption. The authors consider collision, overhearing, control packet overhead, and idle listening as the main sources of energy waste. Also, the work considers the energy consumed by a transceiver to switch between operating modes (active and sleep) which could be greater than that needed to stay always active. Experiments are conducted using two different radio transceivers to evaluate the proposed scheme. In experiments, a theoretic upper bound on channel utilization is used as a performance measure. The results show that the proposed El-MAC protocol is directly applicable to TDMA scheduling approaches. 
%-------------------------------------------------------------------
\subsection{Hybrid MACs}
\label{subsec:hybrid_MAC}

In \cite{wang2010cross}, the authors investigated QoS-based routing in the MAC layer. The paper exhibits a new cross-layer effective protocol of MAC, referred to as Hybrid MAC (HMAC), acceptable for WSNs relating to latency, the efficiency of energy, and design complexity.  HMAC allows the awareness of compromise between different performance measurements as it couples Time Division Multiple Access (TDMA) based MAC protocols and channel allocation schemes from existing contention-based. Compared with other TDMA-based MAC protocols, HMAC presents an extremely low-cost solution as it uses a new wake-up scheme, every node does not have to keep neighborhood information and a little slotted frame structure. HMAC uses three measures to evaluate its performance:
\begin{itemize}
    \item Average energy consumption;
    \item Average end-to-end delay;
    \item Delivery ratio.
\end{itemize}

Experiments show that for energy consumption the nodes in HMAC consume less energy and the remaining energies of the nodes in HMAC are distributed evenly than the existing MAC protocols. For the end-to-end latency, HMAC significantly reduces the end-to-end delivery latency and provides higher delivery ratios due to the use of a little frame length that helps nodes to fast move packets to their next hops and, hence, reduces many collisions and queuing delays.

\begin{figure}[t]
    \centering
    \includegraphics[width=0.95\linewidth]{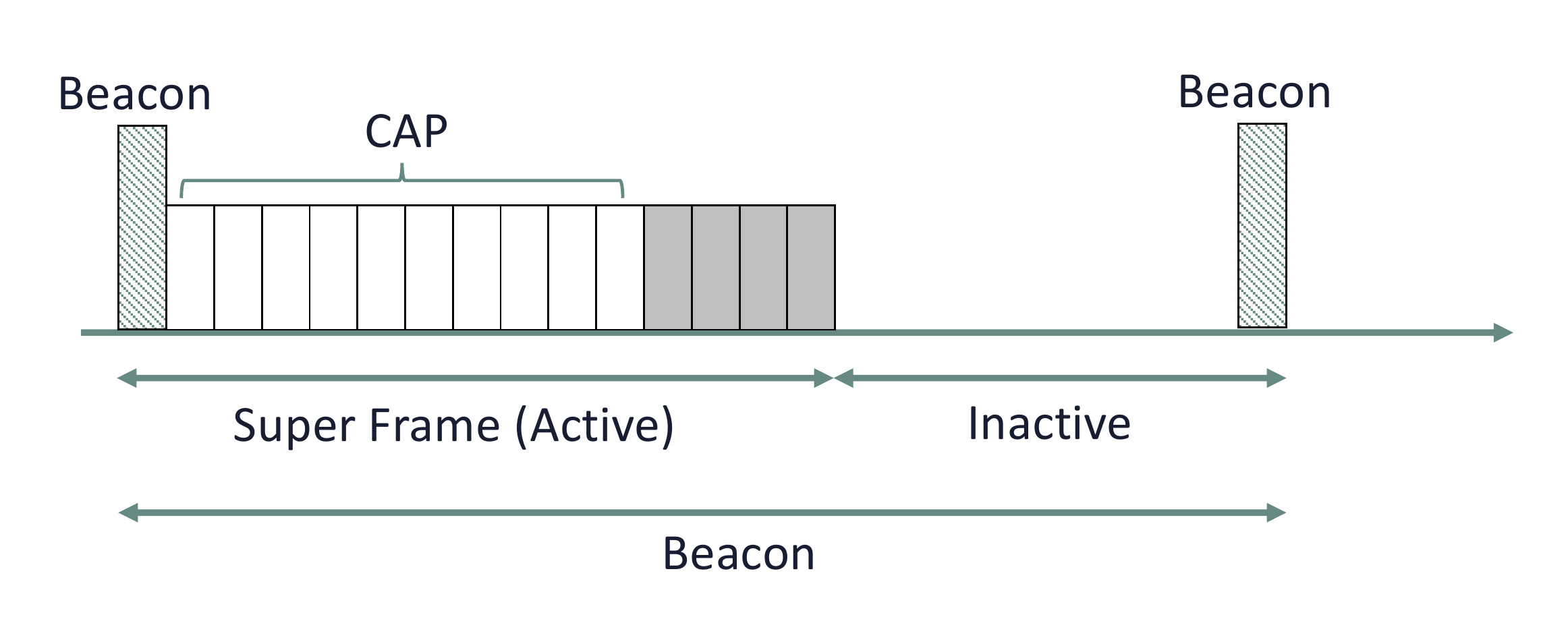}
    \caption{Enabled mode beacon for the structure of superframe.}
    \label{fig:beacon_superframe}
\end{figure}

In \cite{gilani2013adaptive} the authors present a CSMA/TDMA Hybrid MAC protocol for improving features of IEEE 802.15.4 standard in terms of energy consumption and throughput. The proposed protocol overcomes the weakness caused by the CSMA/CA method, utilized in IEEE 802.15.4 standard, in conditions like high loads. In CSMA/CA, the collision avoidance mechanism is not efficient in the case of a large-scale WSN. Also, the slotted CSMA/CA used in IEEE 802.15.4 causes little network throughput because of the collisions that result from many concurrent transmissions, at the beginning of a new superframe, see Fig. \ref{fig:beacon_superframe}. Regarding these issues, the work proposed in \cite{gilani2013adaptive} allows the coordinator to suitably divide the Contention Access Period (CAP) between slotted CSMA/CA and TDMA relative to the collision level detected on the network and the data queue state of nodes. Accurate information of queue state is determined using reserved bits in the standard data packet header. Two main changes are applied to standard IEEE 802.15.4, the first change is applied to add a TDMA period to IEEE 802.15.4 standard using the beacon frame data fields and the other is applied to enable the coordinator to change the border between CSMA and TDMA in CAP. Experiments are conducted on \textit{OMNeT++}. Experimental results of the hybrid protocol are examined to IEEE 802.15.4 standard. Two different conditions have been tested, namely (BO, SO)=(7, 6) and (BO, SO)=(3, 2). The results show that the CSMA/TDMA hybrid protocol enhances the throughput of the network by 1.6 and 2.3 times bigger than the standard IEEE 802.15.4. The average energy consumption has been reduced by 38\% and 70\% for the two conditions of (BO, SO)=(3, 2) and (BO, SO)=(7, 6), respectively.

%----------------------------------------------------
\subsection{Adaptive/Dynamic MACs}
\label{subsec:adaptive_MAC}

Integrating wireless sensors with mobile phones confront environments with dynamic nature where both requirements of the application and wireless conditions environment are changing quickly. Existing MAC protocols cannot provide optimal performance for such dynamic environments. Pursuing a one-MAC-fits-all approach avoids applications from achieving the changing environmental conditions, needs of different workloads, or Quality of Service (QoS) requirements. 

The authors in \cite{ sha2013self } present a Self-Adapting MAC Layer (SAML) which selects the most appropriate MAC protocol for the current conditions and requirements. SAML supports run-time switching among different MACs. As it is equipped with a MAC Selection Engine that enhances the dimensions of latency, energy consumption, and reliability, and considers dynamic external interference.  The engine includes three major modules, traffic monitor, noise monitor, and classifier. The application traffic pattern had been kept on track by the traffic monitor by calculating the mean and variance of Inter-packet Interval (IPI) of the send commands called by the application. The noise monitor calculates the external interference degree in the environment by calculating the mean and variance of the Received Signal Strength (RSS). The Classifier is finally used to determine the best MAC relative to the existing application-specified REL order, i.e. Reliability (R), Energy consumption (E), and Latency (L), and the values released from the Noise and Traffic Monitors. Experiments are conducted on TinyOS 2.x.  Experiments using three models containing a maximum of five MACs. For evaluation, a new gateway device that integrates an IEEE 802.15.4 radio with Android phones is used. The results show the efficiency of SAML in avoiding memory bloat as well as its efficiency in controlling the network nodes.

The work in \cite{qiao2018intelligent} introduces a MAC protocol selection model for the dynamic network environment. Various network parameters are chosen to construct the feature data set to train a classification model. When a satisfying classification accuracy is reached, the optimal MAC protocol could be selected by the MAC selection module. A two-stage selection model is proposed. Stage one is the classification learning process; obtaining an optimal classification result using the Sequential Minimal Optimization (SMO) algorithm which trained different types of feature datasets. Stage two is the selection decision process; the network nodes, with the help of a classifier model, choose an appropriate MAC protocol. The CSMA/CA and Dynamic TDMA are the candidate MAC protocols. Simulations are conducted to evaluate the performance of the selection module. The simulation results show that the proposed model achieves a weighted average Probability of Correct Classification (PoCC) of 94\%. This is the best PoCC ratio compared to those achieved by other benchmark  models such as Naive Bayes, J48, and Random Forest (RF). Moreover, the results show that the MAC protocol selection model had the capability to select the appropriate MAC protocol that best fits the current network environment. 

In \cite{mastronarde2016reinforcement}, the authors present a rate adaptive fully distributed CSMA/CA protocol. By giving users reduced congestion windows, the proposed protocol gives access probability to users who need to send critical data. It assists users in minimizing their energy consumption depending on their limitations of delay by collaborating with the rate-adaption algorithm at the physical layer. In particular, a multi-user delay-sensitive energy-efficient scheduling problem is formulated as a Markov Decision Process (MDP). An RL algorithm is proposed so the single-user problems will be solved online. This solution enables decreasing energy exhaustion to users relative to their delay constraints, bear in mind that traffic, channel, and multi-user dynamics are not known beforehand. Compared to the conventional CSMA/CA protocol, the rate-adaptive CSMA/CA protocol proposed in \cite{mastronarde2016reinforcement} determines congestion windows (CWs) to increase the probability of channel access for users who want higher transmission rates.  In that way, the MAC protocol serves the goal of decreasing the consumption of energy for every user based on their delay constraints when linked with the suggested algorithm of transmission scheduling. Moreover, using this approach, when a user takes the channel, other users stop their counters (backoff). Once the user frees the channel, other users resume their counters assuming that other users reset their counters (backoff) at the end of every time slot. Accordingly, a user’s counter (backoff) will be updated to show its present state. Experiments are conducted to evaluate the energy and delay performance of the proposed protocol against the conventional CSMA/CA through \textit{MATLAB} simulations. The results show that the rate-adaptive CSMA/CA protocol allows all users to adhere to their constraints and uses 65\%  less power than the traditional protocol of CSMA/CA.

\begin{figure}
    \centering
    \includegraphics[width=8cm]{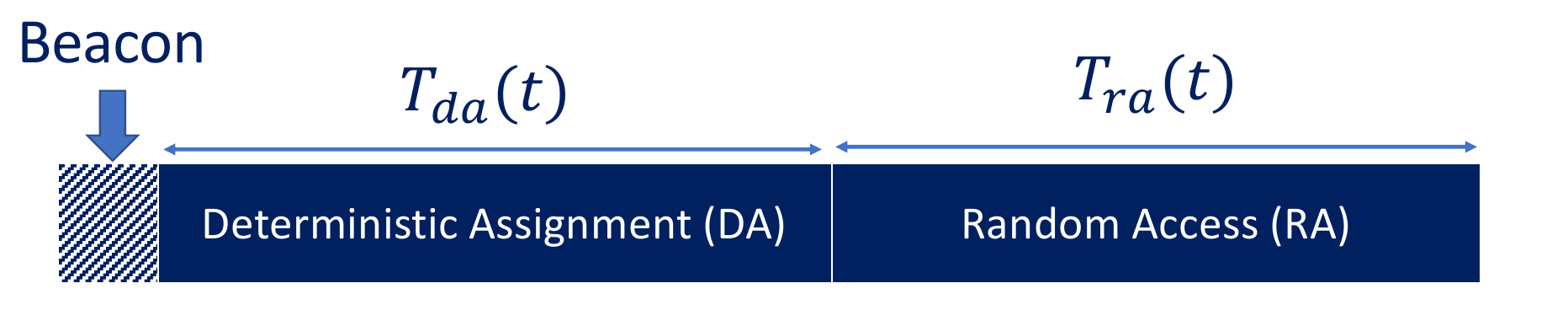}
    \caption{Structure of the frame for Virtualized Wireless Networks}
    \label{fig:Frame_strucure_VN}
    \end{figure}

The authors in \cite{shoaei2019reconfigurable} present a learning-based MAC which is reconfigurable where the division between contention-based and contention-free access regimes in every frame is changed according to the status of the network. The logic behind the suggested MAC protocol allocates devices that have a high probability of transmitting packets to the disagreement-free regime, while the remaining devices are permitted to engage in competition in the disagreement-based
regime. This scheduling is formulated as an optimization problem to maximize the network throughput while maintaining the slice reservations. The proposed protocol employs an RL algorithm based on Thompson Sampling (TS) to collect arrival packet probabilities of devices. Based on learned probabilities, an algorithm is developed in which the Deterministic Assignment (DA) and Random Access (RA) partitions are determined by a threshold. Devices with expected throughput higher than a certain threshold are acknowledged for DA, while the remaining devices transmit in the RA regime. Experiments are conducted to evaluate the proposed scheme in terms of throughput and delay. Experimental results prove that the proposed scalable MAC outperforms other schemes.
The authors chose \textit{airtime} as a slicing metric for the virtualized wireless network since it guarantees isolation among slices. Time is divided into fixed frames each with index $t$ and it is assigned by the AP using the DA regime and RA regime for the periods of time given by $T_{DA} (t) \leq T_{max}$ and $T_f - T_{DA} (t)$, respectively. The detailed illustration for the frame structure is given in Fig. \ref{fig:Frame_strucure_VN}.

%Wireless sensor networks (WSNs) have become increasingly popular and have proven beneficial in a variety of applications. A WSN typically consists of small, inexpensive, and low-power devices that provide data acquisition, processing, and wireless communication functions. One of the most necessary characteristics of a WSN is the endurance of the network. Sensor nodes can be deployed in remote areas such that replacing or maintaining batteries is a difficult or impossible task. Access to the shared communication channel is the responsibility of the Medium Access Control (MAC) protocol, also it is intended to ensure that successful radio management happens with efficient energy management. By avoiding the waste of energy sources, i.e. Overhearing, collisions of packets, inactive listening, enormous,i.e. excessive, retransmissions, and efficient energy management can be achieved.  Adaptive behavior that considers the actual network conditions is urgently needed to address the network lifetime issue.

The work in \cite{galzarano2013ql} presents a new contention-based MAC protocol for WSNs referred to as, QL-MAC. The proposed protocol employs the Q-learning algorithm to reduce energy consumption based on the actual network load in the neighborhood by finding an efficient wake-up strategy. It benefits from a cross-layer interaction with the network layer to enable a better understanding of communication patterns and thus reduces energy consumption due to idle listening and overhearing. The goal of QL-MAC is to permit the nodes inferring each other's behavior to accept a good active/sleep policy scheduling. Q-learning is the most popular and powerful algorithm under the umbrella of RL. Each node must decide whether it should be in sleep or in active mode during each individual time slot. Each node stores a set of Q-values associated with a particular slot inside the frame. The Q-value provides an indication of the benefits derived by a node from being awake throughout the associated time slot. The updates of the Q-value through time are based on a number of specified  events happening through every frame at the same slot. Simulations are conducted using \textit{OMNET++} to compare the performance of the proposed protocol against related MAC protocols in terms of nodes' energy average energy consumption and the Packet Delivery Ratio (PDR). Results show that QL-MAC outperforms other MAC protocols as it allows nodes to save considerable energy due to the sleep/wake-up radio schedule. Also, the results show that when the number of slots declines, generally QL-MAC exhibits improved performance relative to the PDR, but the energy spent by the node tends to increase.

%--------------------------------------------------------
\subsection{ALOHA-based}
\label{subsec:ALOHA_based}

The work in \cite{chu2012aloha} presents a Q-Learning-based MAC with informed receiving for WSNs named ALOHA-QIR. ALOHA-QIR has the advantage of low computation, overheads, and simplicity of ALOHA-based scheme. In addition, Q-Learning, as an RL algorithm, is selected to avoid collisions and retransmissions facing ALOHA-based schemes. Q-Learning stateless is employed in ALOHA-QIR to gain experience in learning. The Q-value that exists in every node is individual for each slot which describes the desirability of the slot selection. At the learning time, the nodes keep jumping to distinct slots to find the perfect one, and they want to listen to keep away from losing any information from the earlier jumps. Informed receiving and ping packets are applied to switch nodes to a rest mode if it could happen when it is important to make the energy efficiency better, exceptionally when traffic operation is low. Experiments are conducted to compare the performance of ALOHA-QIR with basic Slotted-ALOHA. Five performance metrics are considered, normalized throughput, end-to-end delay, the energy cost per bit throughput (indicates the average cost of bringing data), energy cost per second (determines the proportion of consumed energy by receiving /transmitting), and the network lifetime (compares the cost of energy by overheads and data). Simulation results confirm that the ALOHA-QIR achieves over double the throughput of basic Slotted-ALOHA. Moreover, QIR-based networks can survive at least 25 times longer than ALOHA–based networks. 

\begin{figure}
    \centering
    \includegraphics[width=0.95\linewidth]{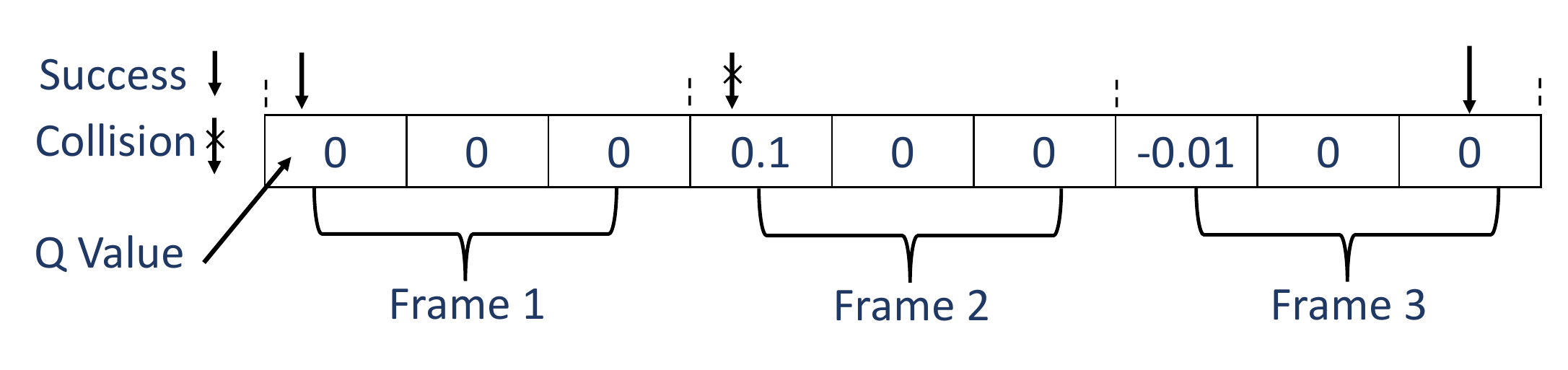}
    \caption{Repeated frames and Q-values examples.}
    \label{fig:repeated_frame}
\end{figure}

Wireless sensor networks (WSNs) exhibit a fast-growing technology with a diverse range of applications for the military, industry, and environment monitoring. The MAC protocol has an important part in maximizing the energy efficiency and throughput of WSNs. Many MAC protocols have been proposed for WSNs that significantly improve energy efficiency and throughput performance. However, these protocols incur higher overheads and exhibit ever-increasing complexity.  Also, Performance evaluation through simulation is common but the reasonableness of recently proposed schemes is questionable. Simpler protocols are needed and can, despite what just has been said, provide efficient energy communication, acceptable throughput, and acceptable delay. RL allows entities to learn efficacious interaction strategies using trial and error in a dynamic environment. 

In \cite{chu2015application}, the authors proposed ALOHA-Q, a protocol that adopts Q-learning with frame-based ALOHA as a strategy of slot selection which provides collision and retransmission avoidance for single-hop networks. In particular, every node has a unique Q-value for each slot in the frame that is updated by failure or successful outcomes of transmission, see Fig. \ref{fig:repeated_frame}. In successful transmission, a reward of +1 is returned otherwise the reward is -1. Slots with higher Q-values are preferred, if multiple slots have the same Q-value, one or more will be randomly selected. When nodes want to transmit, they will wake up, also they will wake up when they have to receive the associated acknowledgments (ACKs). In the experiments, a Markov model of the ALOHA Q-learning is built to analyze its behavior of convergence. Convergence time collected from the Markov model is compared to other related MAC protocols, i.e. S-MAC and Z-MAC. The results of a single-hop network type show that the proposed ALOHA-Q outperforms existing related protocols and provides fast convergence to conditions of steady state, comparable throughput, and delay with better energy efficiency. Also, ALOHA-Q is much simpler.
%################################################
%                Wireless Nets
%################################################
\section{Wireless Nets}
\label{sec:wireless_nets}

The emergence of new wireless communication systems results in a huge increase in the resources of radio demand. The problem of spectrum shortage has been raised due to insufficient spectrum allocation, ineffective frequency band exploitation, and many neglected spectra. Recent studies promote the development of dynamic spectrum access techniques as a promising solution for the spectrum shortage problem.  Several works proposed Q-learning-based Model-dependent solutions to solve the spectrum access in dynamic environments. However, these solutions cannot be effectively adopted in general for handling more complex real-environment.
Recently, DRL has attracted much attention owing to its powerful learning and computation ability. DRL combines Q-learning and neural networks to overcome Q-learning limitations for processing large-scale models.

The work in \cite{xu2020multi} investigates the use of dynamic spectrum access to make the most of multi-channel wireless networks. It assumes the contribution of $N$ users on $K$ channels, with each user having a choice of channel for transmission. Every user chooses a channel and sends a packet with a specific chance of succeeding. Every user who has attempted to transmit a packet, after the regular time slot gets a binary observation indicating the outcome of their sending. The optimal strategy for spectrum access is, in general, computationally expensive. An algorithm for dynamic spectrum access based on partially observable DRL is developed. Experimental results show that the proposed algorithm can improve channel usage by 90\%, without online coordination, message exchange, or carrier sensing. 

\begin{figure}[t]
\centerline{\includegraphics[width=0.45\textwidth]{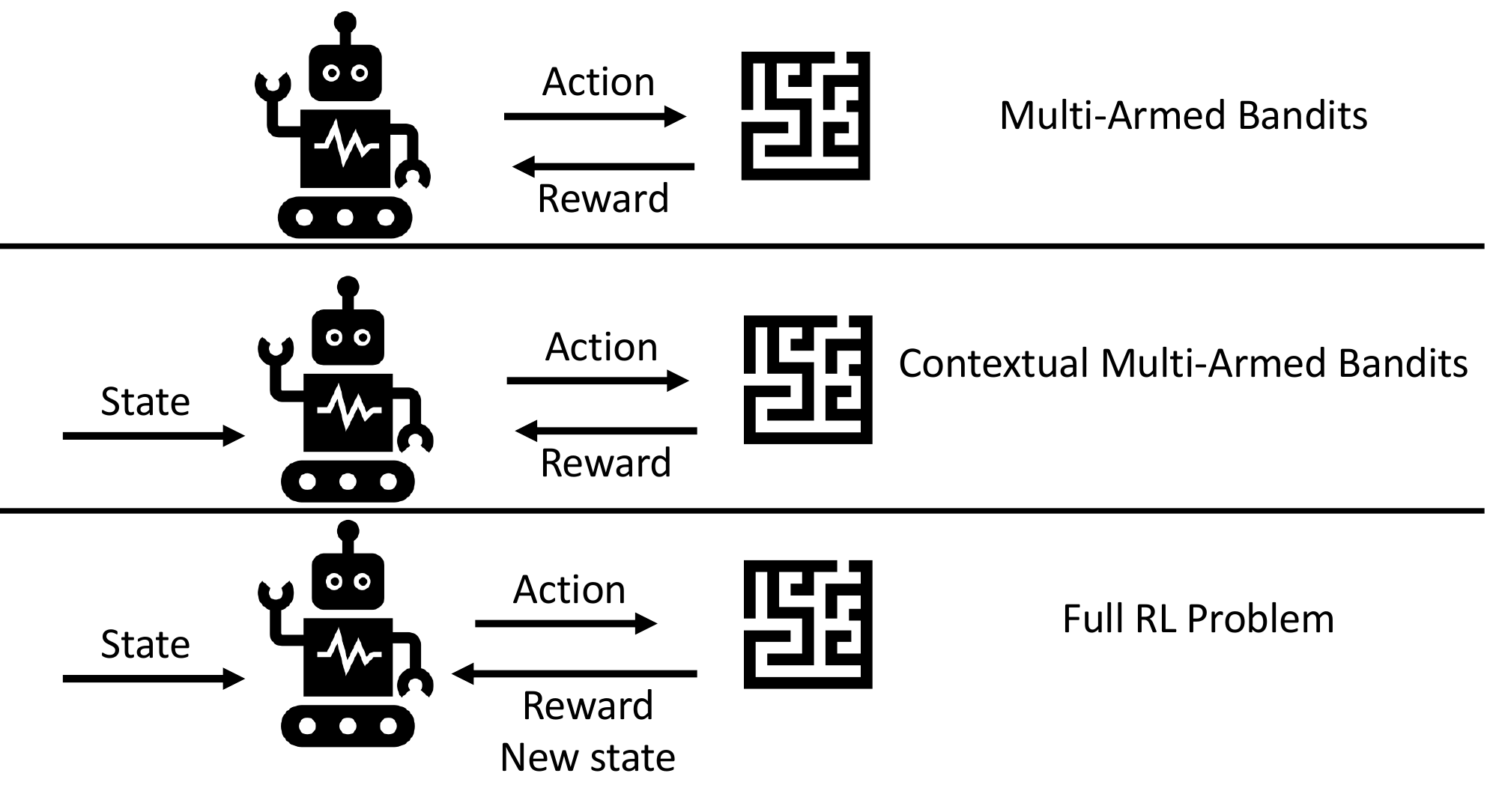}}
\caption{A description for the difference between Multi-Armed Bandits (MAB), Contextual Multi-Armed Bandits (CMAB), and RL.}
\label{fig:MAB}
\end{figure}

Improving resource utilization across networks is important because of the unexpected increase in services and traffic. Wireless network virtualization is one of the most promising solutions as it enables physical resource sharing between different Service Providers (SPs). To achieve higher utilization and lower implementation costs, virtualizing resources into different slices, each for an SP, should be efficiently reached. To support the isolation between SPs, the Quality of Service (QoS) constraint for every slice should be satisfied under all conditions. One way is exact isolation, where different slices can be kept totally isolated by adopting a Time Division Multiple Access (TDMA) scheme. However, this allocation could result in resource underutilization because the slice's reserved timeshare could be partially utilized if there are no active users in each slice. Another way is to use random access protocols such as CSMA/CA to progressively manage the time sharing of every slice depending upon the number of active users. That’s why they can defeat the usage problem. On the other hand, they suffer from isolation problems due to inevitable collisions coupling from different slice flows. These major problems with TDMA and CSMA protocols based encourage the development of a new multiple-access protocol to ensure the required QoS conditions for every slice plus the resource utilization. 
The work in \cite{shoaei2015learning} introduces a hybrid adaptive protocol of TDMA-CSMA MAC to ensure network isolation between Service Providers (SPs) in a virtualized network. The authors in \cite{shoaei2015learning} schedule the AP of high-traffic users from many slices using TDMA while low-traffic users contend for transmission in the CSMA phase. That is why the proposed approach can benefit from the opportunistic nature of CSMA and the isolation advantage of TDMA. The main aim of this approach is to provide isolation between the slices over every superframe. Specifically, it tries to decrease the effects of actions in one slice on other slices. If the traffic arrival statistics are not available, this scheduling will be built as a Multi-Armed Bandit (MAB) model, in which every arm belongs to possible scheduling. Because of arms dependency, modern policies are not directly applicable to this problem. That’s why an index-based policy is presented where decisions and updates are depending on learning indices given to every user instead of every arm.  In order for indices to be updated, observations from the CSMA phase are integrated with TDMA information to add a new phase of exploration for the MAB problem. See Fig. \ref{fig:MAB}.
Simulations are conducted to evaluate the proposed protocol. The results confirm the efficiency of the isolation policy between slices. In addition, the proposed protocol increases the system throughput.

Contention-based protocols are the most common MAC layer access plans in today's wireless networks. In low-load conditions, it has been proven that contention-based protocols perform better than resource-allocation techniques such as TDMA and FDMA. However, with higher traffic loads, more collisions tend to be created by random access techniques, and as a result, an urgent need arose for scheduling protocols. Distributed Coordination Function (DCF) protocol to reach multiple access communication for decentralized wireless networks uses CSMA/CA. It uses a mechanism that exponentially backs off for decreasing MAC layer collisions in the IEEE 802.11 standards. Several works in the literature have shown that this exponential back-off mechanism and its enhanced variants are not efficient under unknown changes such as packet arrivals and user exit/entry.
The work in \cite{amuru2015send} studies the problem of back-off window optimization from an RL perspective and proposes algorithms of online learning that know the back-off optimal plans under unknown changes. In the proposed protocol, the RTS-CTS handshake mechanism is represented via an MDP model. The proposed protocol chooses the back-off window depending upon the optimal state of the system instead of the exponential mechanism which is introduced in the IEEE 802.11 standards. A learning algorithm such as Post Decision State (PDS) is proposed to increase the speed of the learning process as it differentiates the unknown and the known components, both probabilities of the transition of state and costs/rewards, to reach a better and faster learning performance. If compared to conventional Q-learning algorithms, the introduced PDS-based learning algorithm adventurous system partial information so that less information needs to be learned. Also, it eliminates the need for exploring actions that normally delay the learning process. In \cite{amuru2015send}, the back-off window optimization problem is addressed in a single-user setting. Experimental results show that for a single user, a smaller back-off window is preferred when the network load is smaller and vice versa.

A new pattern called Wireless Mesh Networks (WMNs) is developed to supply extensive coverage of the network without infrastructure usage. In these networks, nodes are used simply as relays to spread data from the source to the destination using multi-hop paths. WMNs usually employ IEEE 802.11 standards whereas the MAC protocol depends on CSMA/CA and  DCF. The physical layer uses many coding and modulation techniques to support various rates. Employing a higher transmission rate needs higher transmission power to reach the required SINR on the receiver. This result in higher interference between the communicating nodes and generally reduce network throughput. Several studies are proposed in the literature to adapt the transmission rate of WMN. However, these studies do not differentiate between packet collisions and channel errors in case of transmission failure. The work in \cite{al2014multi} presents a new RL to adapt the rate of transmission for the dynamic conditions of WMNs referred to as (RARE). The proposed algorithm uses the access probability of the wireless channel to determine if the transmission rate will be updated or not. It alleviates the bad impact of unnecessarily updating the transmission rate  when the transmission failure happens due to channel errors instead of interference. The proposed, Q-learning-based, RARE is an agent-based algorithm where each node computes the access probability of the medium-based communication on the number of unsuccessful transmissions and the current transmission rate. Also, every node periodically collects a \textit{Hello} message from its neighbors containing a traffic-load estimation, the channel access probability, and the transmission rate. RL is used at every node to learn from earlier actions whether it is important to update the transmission rate. This, in turn, decreases the negative consequences of acquainting the transmission rate when the throughput degradation happens by channel errors  not by interference. Experiments are conducted to compare the performance of the RARE algorithm against other benchmark schemes.  In experiments, Fisher’s Least Significant Difference (LSD) is used as a performance metric to compare the throughput obtained by various methods. The simulation results show that RARE achieves higher throughput compared to the state-of-the-art algorithms for different transmission loads and the number of contending nodes.

\begin{figure}[t]
    \centering
    \includegraphics[width=0.95\linewidth]{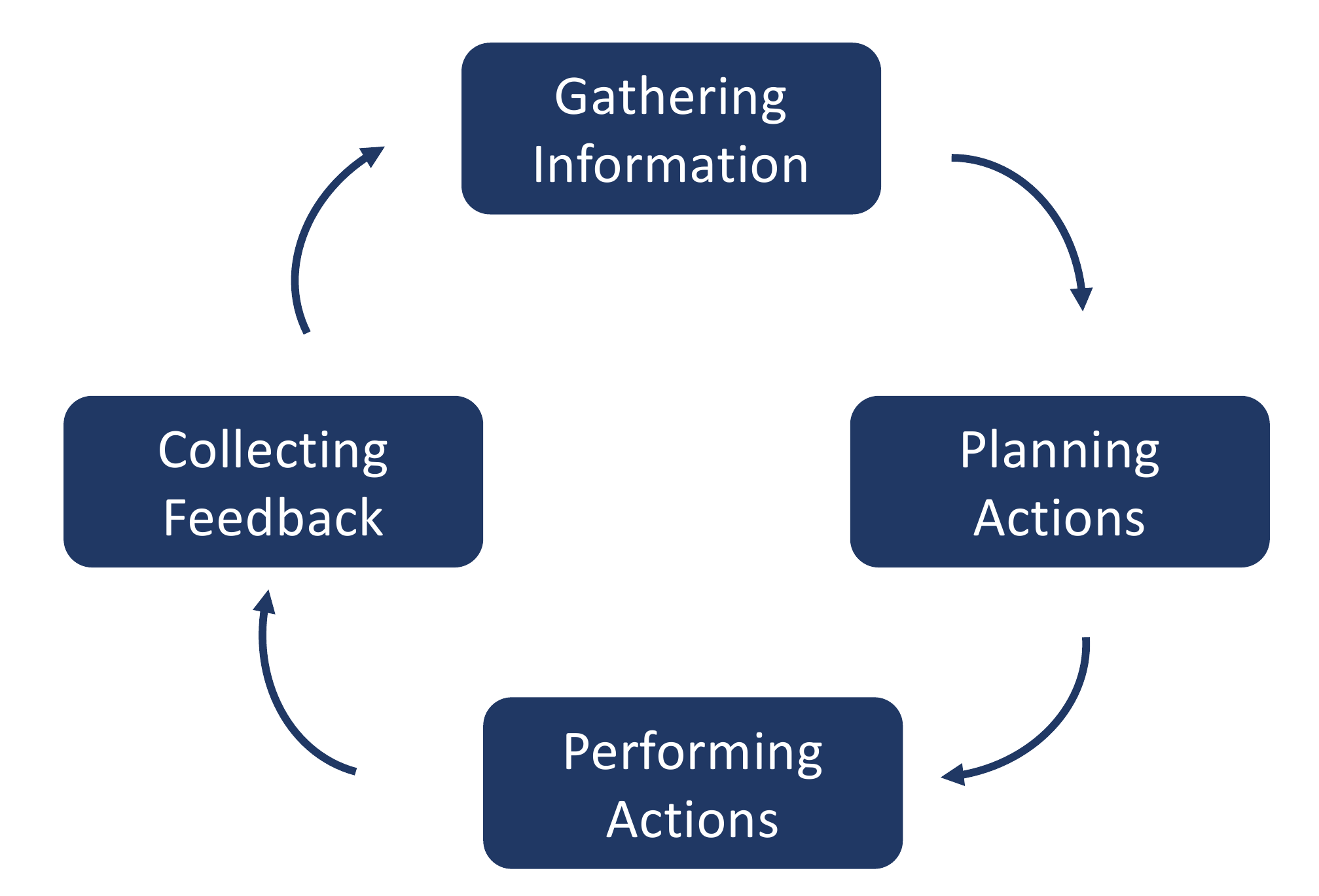}
    \caption{A general view of the four levels of cognition cycle: first, information gathering, second, actions planning, third, actions, fourth, feedback collection.}
    \label{fig:four_levels_cog}
\end{figure}

The proliferation of wireless communication technologies has resulted in an increase in heterogeneous, co-located wireless networks with different requirements, mobility capabilities, data rates, and coverage. Run-time cooperation between devices from heterogeneous wireless networks becomes a challenge. One way to support connectivity between co-located devices is to manually group them into different sub-nets, according to their communication technology. This approach is quite complex and inefficient. Direct cooperation between the independent networks through shared resources such as intermediary nodes for routing can address these problems and improve network performance. However, the configuration problem is further complicated, as management deals with many heterogeneous networks, distinguished by differing requirements and capabilities. Using a cognitive entity to initiate and supervise the entire cooperation process is the most promising solution for the above-mentioned problems.
An RL technique, Least Square Policy Iteration (LSPI), had been proposed in \cite{rovcanin2014reinforcement}, for high-level network optimization in heterogeneous co-located networks. LSPI is a form of ML that collects knowledge by using a series of trials and errors, that represent the network characteristics, to evaluate the impact that different service combinations have on the requirements of each network. Different decision-making policies are not required by the proposed cognitive engine since it constantly selects environment samples and updates the sampling matrices, see Fig. \ref{fig:four_levels_cog}. Therefore, each time a new sample is collected, LSTDQ is reinitiated using the same policy. Compared to other work, the proposed approach does not require prior knowledge of service impacts on network performance. Experiments are conducted to evaluate the performance of the proposed cognitive engine against a linear programming-based reasoning engine. Unlike other works, the proposed LSPI is capable of adapting to dynamic network conditions and learning optimal network configurations.

For next-generation wireless networks, efficient video streaming will play a key role in various important applications. Unfortunately, current mobile Internet architectures are unable to achieve Quality of Experience (QoE) constraints for the acquired video, e.g. small zapping time, playout quality, or limited delay of delivery. This is attributed to the limited radio resources, the dynamic channel characteristics, and the dynamic characteristics of video content. To address the aforementioned problems, several works have been proposed. In the majority of these works, the control process uses the receiver buffer and channel states. With this being said, this information can arrive at the controller with a certain delay in real networks. One method to address the issue of time-varying properties of both the channel and the encoded videos is to know and update the policy of optimal layer filtering online. RL methods are well appropriate to regularly update the policy and the function of the state value.
The authors in \cite{changuel2012online} introduce a cross-layer control mechanism and scalable video stream to mobile receivers. The goal of the proposed approach is to maximize the quality of the received video while taking into account the variations in the characteristics of both the transmitted content and the channel. To deal with the dynamic characteristics of multimedia content and wireless channels, RL methods are used to dynamically update optimal policy. The proposed mechanism focuses on different classes of online RL algorithms, such as Temporal Difference (TD), which targets to directly estimate the function of action value, i.e. Q-function. Several experiments have been conducted to evaluate the proposed layer filtering using both standard and real video sequences. In experiments, the absence and delay of channel state information are considered. PSNR is considered as the quality metric for evaluation. Results show that with the absence or delayed channel state information, the performance of the proposed solution is slightly degraded. Since PSNR is not enough to reflect the perceived video quality, subjective quality metrics should be considered and estimated at the transmitter. 

\begin{figure}[t]
    \centering
    \includegraphics[width=0.95\linewidth]{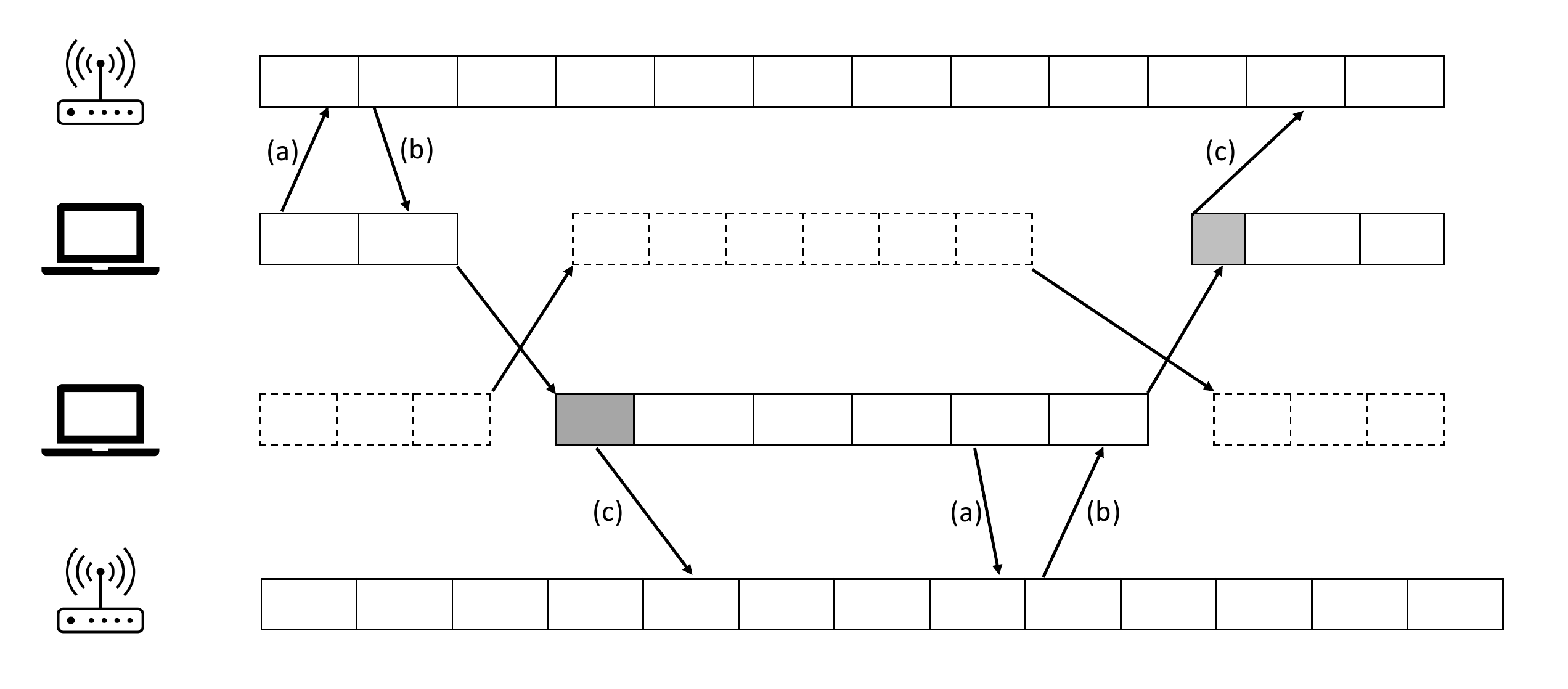}
    \caption{Subnet 1 and subnet 2 with operation of MF-WT}
    \label{fig:subnets}
\end{figure}

From a control point of view, we can divide the MAC protocols into two categories: distributed and centralized (i.e. infrastructure-based) MAC protocols. Distributed MAC protocols achieve lower throughput compared with centralized protocols, but multi-hop transmission is not provided by centralized MAC protocols. This could be attributed to the hidden terminal problem arising from the discrepancy between subnets. Providing multiple channels or frequencies for transmission is an efficient method for centralized MAC protocols to solve the hidden terminal problem between subnets. A small number of investigations on systems of multi-channel concern the architecture of multi-channel for transmission of multi-hop in centralized wireless networks. These works rely on registering a new Wireless Terminal (WT) and employing a large number of channels. In this way, the implementation cost of these protocols, and the consumption power of every WT are high. Multiple Frequency Forwarder Wireless Terminals (MF-WTs) is proposed in \cite{cheng2010adaptive} to enable multi-hop transmissions for centralized MAC protocols. An MF-WT is designed to join two or more overlaying asynchronous subnets by changing these subnets' frequency, see Fig. \ref{fig:subnets}. Despite that, many topologies of the network have no MF-WT available, because the MF-WT must exist in the same area with overlaying subnets. The authors in \cite{cheng2010adaptive} suggest a mechanism of multi-hop referred to as Adaptive Channel Switching (ACS) for centralized MAC protocols. ACS  mechanism provides efficient bandwidth utilization by preventing channel divisions between centralized protocols subnets. It allows subnet transmission using multi-hop and overcomes the problem of the hidden terminal by dividing the whole bandwidth into three channels: 
\begin{itemize}
    \item Relay Channel (R-channel) 
    \item Data Channel (D-channel)
    \item Control Channel (C-channel)
\end{itemize}
The R-channel is used to relay packets to adjacent subnets by subnet boundary stations, while D-channel is used for data packet transmissions, and C-channel is for control-signals exchange. The proposed approach provides two distinct operation modes, 1) Free Mode (F-mode) and 2) Restricted Mode (R-Mode). In F-mode, stations are allowed to use the C-channel to send control signals and use the D-channel to send data packets, they also use the R-channel to communicate with other stations in R-mode. Stations in R-mode are restrained.  It uses just the R-channel to send data packets based on the CSMA/CA and RTS/CTS mechanisms.

In literature, a great attention had been given to the problem of collaboration in wireless networks. As effective collaboration among nodes undoubtedly leads to a great improvement in the performance of the wireless network. Different studies have been presented and shown that the problem of collaboration can be represented as a Markov Decision Process (MDP) framework. In these studies, each node has a local MDP comprising its own reward functions, state, and action. The wireless nodes then collaborate to find a near-optimal solution by dealing with the limited information with their neighbors. The studies considered the use of distributed MDP plans for the problem of collaboration. Although, they do not look into the methods of distributed learning. The research in  \cite{naddafzadeh2010distributed} presents an MDP architecture for adapting the transmission power and probabilities. This adaptation happens in the relay and source nodes in order to fulfill the network's highest throughput per unit of energy consumption. The research, also, introduces a new learning method named the Distributed Reward and Value function (DRV). It is a conjunction of Global Reward-based Learning (GRL) and Distributed Value Function (DVF) methods. In the proposed (DRV) method, to maintain a balance between the long and short-system-term rewards, both value functions and rewards are shared among neighbors. In particular, the quick reward stresses the present system state, and the value function is a strategic average of the rewards. Thus, by delivering both reward and value functions at the nodes, a complete view of the system is obtained. Simulations are conducted to evaluate the proposed learning methods against other learning methods and noncooperative schemes. Simulation results show that all the learning approaches greatly surpass the not cooperative design by providing a minimum improvement of 50\%. Particularly, learning methods can exploit the dynamics of channels, and accordingly adapt their probabilities of collaboration, to accomplish a higher throughput per unit of consumed energy and lower network collision.

%--------------------------------------------------------
\subsection{Mobile Ad-hoc Networks (MANETs)}
\label{subsec:MANETS}

%The performance of prevalent wireless MAC protocols is affected by the network contention level and the capabilities of the underlying network nodes. Several wireless MAC protocols have been proposed, ranging from contention-based to slot-based protocols. While contention-based MAC protocols such as CSMA suffer from high performance degradation under high levels of contention, slot-based MAC protocols such as TDMA behave in a contrary way. Adapting a hybrid behavior between TDMA and CSMA received considerable attention.  Sub-section (Mobile Ad-hoc Networks (MANETs)) will explain that subject in more detail.

The work in \cite{hu2011load} extends the prior efforts by dynamically weighing the pros and cons of CSMA and TDMA for MANET. The paper introduces a hybrid CSMA-TDMA MAC protocol called Load-Adaptive MAC (LA-MAC). It is specifically planned and executed on MANETs formed by MIMO-equipped USRP nodes. By dynamically switching its operating mode, the proposed protocol is aimed to conduct similarly to CSMA under low collision conditions and TDMA under high collision conditions. Particularly, the LA-MAC protocol offers better performance using a cross-layer PHY-MAC design to command the mode-switching behavior of the protocol based on real collisions as opposed to MAC frame corruption associated with wireless channel effects such as fading. Experiments are conducted on a MANET testbed and two different metrics are used to investigate the LA-MAC performance, Maximum Achievable Data Throughput (MADT) and Round Trip Time (RTT). The results show that, in contrast to other existing protocols, the throughput of LA-CSMA is almost independent of the number of transmitters. In addition, every time slot can be used for transmission. The results also show that the average latency of LA-CSMA is almost 10\% lower than other protocols.

In \cite{cordeiro2018fs}, Current wireless networks are very dynamic and support numerous applications. As an example, a cellular network carries a variety of traffic such as multimedia messages, text, webpage requests, video calls, and voice calls. Each of these traffics has its own requirements and characteristics. Furthermore, it is necessary that the network adapts its attitude based on the application requirements. On the other hand, the requirement for adaptability also occurs in the sublayer of MAC, because it is not possible for an individual protocol to achieve orthogonal requirements such as power consumption, latency, availability, bandwidth, and security. Furthermore, devices and networks should adjust to the context of communication in order to utilize resources usages. The work in \cite{cordeiro2018fs} introduces a highly adaptable protocol platform for the MAC sublayer, called FS-MAC. FS-MAC dynamically switches the used MAC protocol in the network, to preserve using the protocol that is more effective for the current network status. The switch protocol depends on a set of fuzzy rules, that can be adapted according to the network administrators and application demands. The proposed architecture is extensible for additional MAC protocols. FS-MAC is not considered to be a new protocol for MAC, rather it authorizes the use of many protocols in the MAC layer. Each of these protocols will be used in the appropriate conditions where it is highest effective. Extensibility is the most intrinsic feature of the proposed platform as a new monitoring metric, that enriches the decision, and new MAC protocols could be added in any later phases. Furthermore, the decision rules could be changed by the network administrator to deal with new applications or to meet the prerequisite of the network. The rules are centralized for the best protocol. In decentralized networks, i.e. Mesh or ad-hoc networks, a leader election algorithm to choose the coordinator can be employed. The proposed framework is composed of three main modules: sensing, decision, and change modules.  
The sensing module assembles data to determine which MAC protocol must be active. The Decision module computes, using fuzzy logic, the adjustability of every protocol to the current network after getting the information from the Sensing module. Simplicity and extensibility are the key features of fuzzy logic over other intelligent approaches. In addition, Fuzzy regulation captures the domain knowledge from domain experts, so the system does not require pre-training, and new rules can be added without touching the existing rules. The change module inspects the need to switch the currently used protocol and allows the system to change the MAC protocol only if the adjustability of the current protocol is less than the adjustability of the finest protocol plus a threshold. 
A testbed has been used to measure the platform's capability to respond to different congestion levels. The load in every station is held constant during the experiments while a variable number of transmitting stations is considered. Results show that FS-MAC has delay values and throughput like those of the finest fixed protocol for every scenario, with only 2\% overhead.

The work in \cite{ju2010scalable} proposes a new Scalable Cognitive Routing Protocol (SCRP) to save routing overhead for mobile ad-hoc networks. The suggested SCRP exploits a flooding protocol with intelligence. A neural network-based approach is approved to make nodes familiar with history. Every node forecast future link status based on previous experience and flood RREQ packets along with the predicted strong links and over-predicted good frequencies. %Different simulations are conducted to evaluate the proposed SCRP protocol against other on-demand routing protocols in terms of the following metrics: Overhead: the number of RREQ packets received by nodes, which dominates the number of route control packets. Throughput: Average rate of successful message delivery measured in Kbits per second, which reflects network performance. The obtained results show that the SCRP routing efficiently scales with on demand, routing protocols in terms of network spectrums, dynamics of the network, and size of the network.  Furthermore, it deteriorates to some extent  network achievement.
%------------------------------------------------------------
\subsection{Machine to Machine (M2M)}
\label{subsec:M2M}

M2M networks are expected to be broadly used in many widespread IoT applications such as power grid systems, transportation systems, and health care. A key characteristic of M2M networks is the massive number of devices and parallel network access attempts from these devices. Contention-based MAC protocols are not convenient to provide scalable, flexible, and structured automatic communication for a dense heterogeneous M2M network. Combining the advantages of contention-based and reservation-based MAC protocols in a hybrid scheme has received considerable interest. An efficient and scalable MAC protocol is essential for M2M networks to enable a huge number of devices to concurrently access the channel. Current wireless M2M networks experience more diverse traffic characteristics than usual networks. To enhance the channel usage for M2M networks, the traffic statistical information might be leveraged to efficiently select a MAC protocol or to configure a certain protocol in response to differing conditions. The traffic statistical information, or parameters, may not be readily available or may need to be gathered over time. Therefore, employing a learning algorithm to acquire traffic statistics is crucial.

In \cite{ liu2013scalable} and \cite{ liu2014design} the authors propose a frame-based Hybrid MAC protocol for M2M networks which combines the advantages of both contention-based and reservation-based protocols. In the proposed approach, frames are divided into two parts: Transmission Only Period (TOP), and the Contention Only Period (COP). COP depends on the CSMA/CA access method while the TOP delivers TDMA type of communications. The devices first compete for TOP transmission slots. Transmission time slots during the TOP will be given to the winning devices. To achieve the optimal balance in each frame between the transmission and contention period, an optimization problem will be developed for throughput maximization. Experiments are conducted to evaluate the proposed Hybrid MAC protocol with the contention-based protocol (i.e., slotted ALOHA) and the reservation-based protocol (i.e., TDMA). The comparison considered various metrics including throughput, service delay, and average transmission. The obtained results indicate that the proposed hybrid protocol successfully reduces collision probabilities and increases the overall device's throughput. On the other hand, Slotted ALOHA has a good performance only in low-loaded conditions, and TDMA has a good performance only in high-loaded conditions. The hybrid protocol achieves a lower delay than Slotted ALOHA and on-the-bar with TDMA.

%The authors in  propose a scalable Hybrid MAC protocol consisting of a contention period and a transmission period for heterogeneous M2M networks. The hybrid protocol introduces a frame-based procedure for the contention and reservation processes of different devices. The frame composes of two portions: COP and TOP. The COP is based on a p-persistent CSMA mechanism which allows different devices to contend transmission slots with their own priorities. Successful pieces of equipment authorized sending data all along TOP.  This way, the data connection type will be granted to TDMA. The incremental conflict priority method is followed to assure access justice among all devices by raising the contention probability for failed devices at the next frame.  To achieve the optimal tradeoff between the contention and transmission period in each frame, an optimization problem is formulated to increase the channel utility. The optimization problem considers the optimal contending probability during COP and the best amount of equipment allowed for communicating all along TOP. Experiments are conducted to evaluate the proposed Hybrid MAC protocol in terms of channel utility, packet drop ratio, average transmission delay, and energy consumption. The results show that the proposed hybrid protocol outperforms the pure p-persistent CSMA and TDMA protocols for the mentioned performance measures.

%-------------------------------------------------------------
\subsection{Cognitive Radio (CR)}
\label{subsec:cognitive_radio}

Cognitive radio (CR) has been raised as a promising solution for the problem of spectrum scarcity in wireless communications. CR improves the efficiency of spectrum usage by allowing CR users to access the gaps in the spectrum. However, higher CR intelligence techniques are still needed, particularly for a number of individual cases of radio environment, to reach the goal of smart radios along with the goal of CR transmissions \cite{ranjha2020urllc}. Wireless technologies are becoming more important and involved in various aspects of our daily life. The increasing number of wireless applications, from satellite control to home appliances, creates problems such as spectrum shortages and a lack of radio resources for those who are most in need. In a Cognitive Radio ad-hoc Network (CRN), the SUs interfere with each other because of the difficulty of synchronization and thus the protocol has to address this issue. In this subsection, we briefly introduce some of the AI-based works in this direction.

The work in \cite{kakalou2015reinforcement} suggests a Multi-Channel Cognitive MAC Protocol for CR wireless ad-hoc networks. The proposed protocol uses an RL model for secondary user (SU) channel selection based on primary user (PU) traffic and with small computational requirements to keep SUs synchronized and avoid collisions that may happen with the PUs. In particular, the authors consider network architecture that has clusters of SU nodes experiencing the same PUs existence. Inside each cluster, local traffic can be exchanged, while inter-cluster communication is achieved by the gateway node. The proposed MAC protocol uses a common per-cluster resource reservation control channel to restrict the collected interference of SUs and to deal with the problems of hidden terminals. The suggested protocol depends on the synchronization reached inside the cluster by using the distributed scheme since the sender and receiver are aware of the existence of the identical PUs and they share identical probabilities of channel selection. Transmissions are identified and SUs are synchronized because PUs and Sus have no collisions between them. Energy acquisition and cyclostationary feature acquisition can be used for spectrum acquisition.
Simulations are conducted using the \textit{OMNET} simulator to evaluate the performance of the proposed protocol against the CREAM-MAC, a protocol introduced in \cite{su2008cream}. For various PU burst lengths, the collected throughput is analyzed. When the bursts are generated every 2s and 0.1s, the proposed protocol is less influenced by the changes in PU burst length as it applies the maximum of the spectrum holes for each PU burst generation rate that learns, and differs from CREAM-MAC as its performance falls substantially by the PU burst length. Simulation results also show that the proposed protocol forecasts and completely evades PUs collisions. %Cognitive radio networks are equipped with wireless devices that are entirely programmable and entrust the network with the OODA cycle.  They make smart decisions by adjusting the characteristics of their physical layer and MAC. 

Based on the way of advertising the FCL between the participating cognitive nodes, Cognitive Radio MAC protocols can be classified as a Global CCC (GCCC) and non-GCCC. The authors in \cite{ shah2011ddh} present a novel Dynamic De-Centralized Hybrid MAC protocol, referred to as “DDH-MAC”, for Cognitive Radio Networks. DDH-MAC protocol lies between GCCC and non-GCCC categories of cognitive radio MAC protocols. It supports many security levels and two selection levels.  The first level of selection lets cognitive nodes learn about deciding which two white spaces to apply BCCH and PCCH. At the moment when the BCCH and PCCH are decided, switches cognitive nodes exchange control information through the empty space, and the selected second level starts. Four levels of security are provided by DDH-MAC, by enciphering the BF using either private or public cryptology designs, the security first level has been achieved and the message integrity is gained using codes of message authentication. Secret exchange of the FCL in PCCH, which is accessible locally and is only acknowledged to the engaging cognitive nodes, presents a second level of security.  An additional security level is accomplished by the addition of a timestamp plus its hash value in the precise data frames transmitted by the CR nodes. Finally, DDH-MAC dynamically adapts new PCCH for every transaction. For evaluation, the time of pre-transmission for scenarios of different DDH-MAC has been calculated and compared with other existing CR MAC protocols. The results show that the time of pre-transmission is 20\% superior on average for DDH-MAC and this also will help DDH-MAC to achieve improved QoS as the nodes do not need to wait. In CR networks, CR users have to incrementally rearrange the properties of their MAC layer in response to the changes of the outside radio settings to reach an improved network performance. Different computational intelligence is required to help CR nodes to decide which MAC protocol should be used.

In \cite {hu2012mac}, the authors consider the MAC protocols as radio specifications and present a modern technique named MAC protocol identification to execute the cognitive cycle for CR. The proposed approach acts in the execution of the reconstruct/adaptive MAC. It enables CR users to verify and ascertain the MAC protocol type and the MAC layer specifications of any network users, involving CR and PUs. Identification results enable the CR user to enhance the spectrum holes fulfillment capability, decrease the interference to reactivate the original users, and promote communications among diverse CR networks. Also, the proposed approach helps CR users to enhance the spectrum sensing frequency/period to decrease power usage. Two experiments are conducted to evaluate the performance of packet length prediction and protocol identification using a support vector machine model. The experiment results show that the proposed approach achieves 92\% accuracy in packet length prediction. While the MAC protocol identification accuracy reaches 90\% with increased PU traffic load.

\begin{figure}
    \centering
    \includegraphics[width=0.95\linewidth]{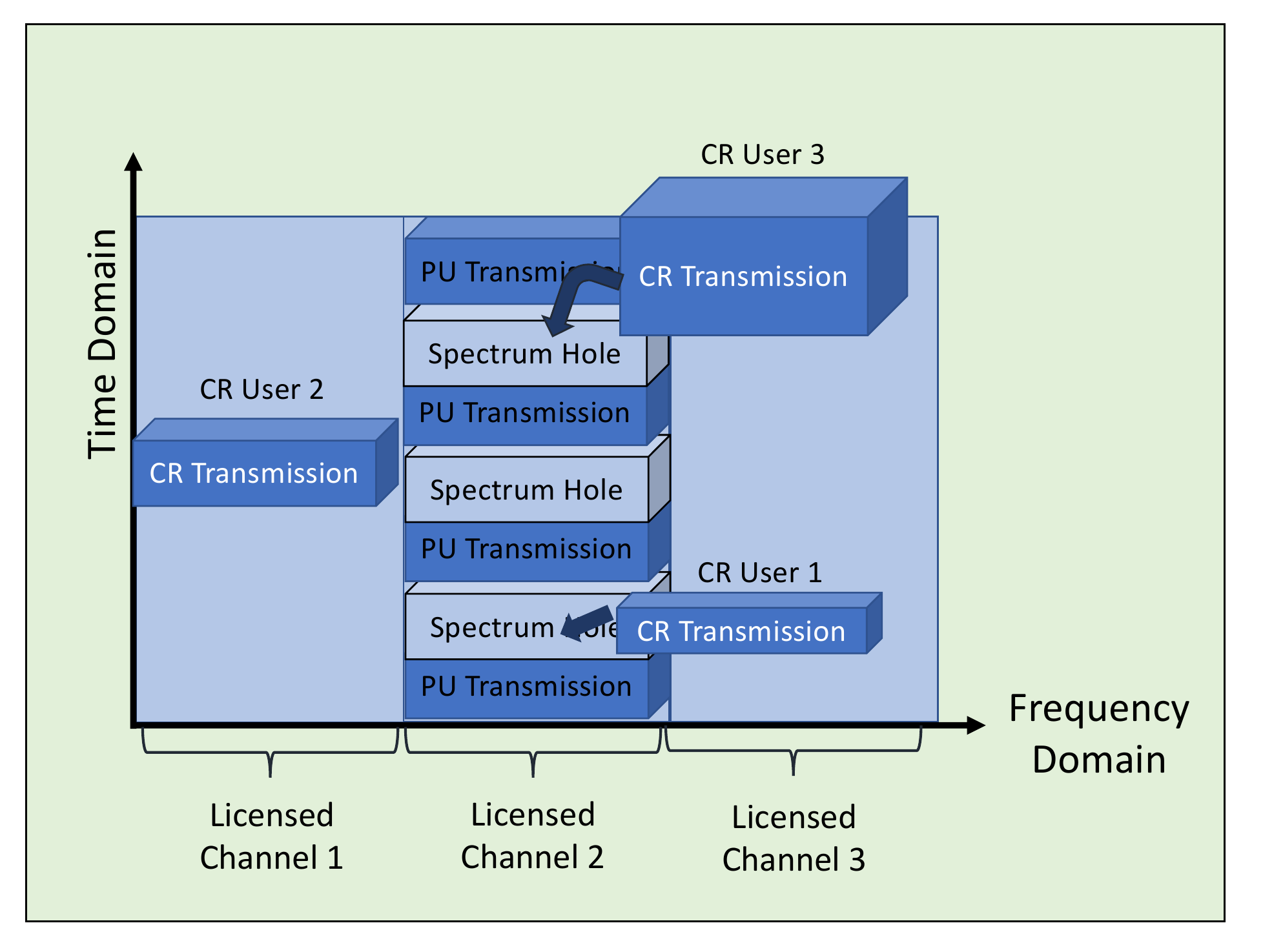}
    \caption{Spectrum holes' allocation of CR transmissions.}
    \label{fig:spectrum_holes}
\end{figure}

The authors in \cite{hu2014mac} investigate recognizing the MAC protocol for applications in cognitive networks considering TDMA, CSMA/CA, and both pure and slotted ALOHA networks. MAC recognition allows CR users to recognize and sense the MAC protocol types for available transmission. Consequently, the identification results enable CR users to adapt their transmission specifications to enhance spectrum exploitation and decrease interference between secondary and primary users. In particular, the received signal strength (RSS) is considered the main feature of CR users. It is used to detect spectrum availability through an SVM classifier. Which in turn increases the detection probability and reduces the false alarms probability. When a CR user attempts to use a spectrum hole, time duration and frequency range are treated as channel access specifications. The frequency range of a spectrum hole is usually determined by spectrum acquisition; however, the time range information is not used. Unlike spectrum capture, MAC protocol identification allows CR users to match their transmission to the timing pattern of a spectrum hole, see Fig. \ref{fig:spectrum_holes}. In literature, research on CR networks suggests that the users in primary networks are non-reactive which is why the primary transmission processes are not influenced by the transmission of CR users. However, in a reactive primary network, the appearance and transmission of CR users will seriously interfere with PU. To address the interference with the responsive PUs, a MAC protocol recognition is required to detect the protocol type of a primary network, and the next coexisting access scheme of CR is applied. Experiments are conducted to evaluate the MAC identification performance considering TDMA, CSMA/CA, pure ALOHA, and slotted ALOHA MAC protocols. Power and time components are obtained and used by SVMs to detect the type of MAC protocol. Each CR user keeps sensing the channel and recording the RSS and channel state durations. Power mean and variance are, in this case, used as power components. The minimum, median, and maximum values of the recorded channel idle and channel busy durations, are used as time features. Experiments show that using the same kernel function the used SVM has different performance for the four MAC protocols. S-ALOHA is easier to identify than the other three protocols as the high contention behavior provides more accurate power features. Experiments also considered the use of different kernel functions for the same MAC protocol. The results show that the linear and polynomial kernels achieve more effectiveness than the radial base kernel. That implies that kernel selection is critical since it influences the identification accuracy.

%##################################################
%          Body sensor Nets
%##################################################
\section{Body Sensor Networks (BSNs)}
\label{sec:body_sensor_net}

Body Sensor Networks (BSNs) are composed of many sensor nodes placed on/around a human body to send and/or receive data through wireless connections. A commonly used low-power BSNs wireless protocol is IEEE 802.15.4 (ZigBee) which supports healthcare applications \cite{gravina2017multi}. Most of these applications use CSMA/CA as the standard MAC scheduling scheme for IEEE 802.15.4. However, CSMA/CA provides unacceptable performance in dense environments with multiple co-located BSNs. Time-slotted channel hopping (TSCH) is a well-suited MAC protocol for healthcare applications as it incorporates access to time-slotted and hopping of multi-channel capabilities to deliver applications with deterministic latency. Also, it improves the reliability of wireless communications in the presence of heterogeneous technologies. In this subsection, we present the recent AI-based MAC work targeting BSN \cite{narsani2022leveraging}. 

In \cite{ngo2017user} a joint sampling plan of dynamic sensor plus a BSNs dynamic MAC scheduling plan based on TSCH has been presented for a healthcare monitoring system. The adaptive MAC scheduler depends on a state machine model that responds to the sensors-generated dynamic healthcare traffic. Additional time slots for the emergency sensors had been allocated automatically by the MAC scheduler. This emergency allocation allows reliable transmissions for high-resolution sensor data. These extra time slots are released back, when the emergency period is expired, to the pool of available time slots of the whole network. Such a mechanism solves the problem of allocating available time slots of TSCH to the required sensors while maintaining the other sensors' QoS. Experiments are conducted to evaluate the adaptive MAC scheduler against other MAC schedulers for TSCH in terms of Packet Delivery Ratio (PDR) and Goodput (G). The results show that the proposed MAC scheduler maintains a great communication channel QoS, nearly 100\% PDR constantly, and obtains a great Goodput ratio in contrast to the other existing MAC schedulers.
%####################################################
%           Spectrum Sensing
%####################################################
\section{Spectrum Sensing}
\label{sec:spectrum_sensing}

Dynamic Spectrum Access (DSA) was introduced to deal with the largely increased need for resources of the finite spectrum. DSA is proposed to open spectrum chunks that are currently allocated for licensed PUs to other unlicensed CR SUs. Existing spectrum sensing approaches for CR DSA are generally categorized into two categories: energy-based detection approaches and signal classification approaches. Apart from the instantaneous spectral and temporal occupancy of the PU, energy-based approaches did not provide much information. This is considered as an advantage to drive the sensing policy, nevertheless, it does not adequately express PU health. Higher-level information about the PU is provided by signal classification approaches, but, in some cases, requires complex and computationally intensive features and supervised training. Following we present some works that explore spectrum sensing for MAC techniques. 

In \cite{ford2017enhanced}, the authors present a learning approach for spectrum acquisition, where the PU application protocol is discovered such that PU knowledge of the application protocol is identified using easy functions for energy detector. The presented approach allows an SU to automatically identify access opportunities only using externally observable energy detector features. This approach includes an unsupervised nonparametric Bayesian Hierarchical Dirichlet Process Hidden Markov Model (HDP-HMM) training method along with efficient HMM recursions for log classification, log state detection, and anomaly detection. Experimental results, on a wireless network testbed, show that hidden states could be learned by HDP-HMM. Hidden states correspond to the states of the major user application layer protocol. HDP-HMM looks for PU traffic abnormality occurring by SU interference. There have been prior attempts for exploiting Deep Reinforcement Learning (DRL) to solve the MAC problems. Compared to RL-based MAC, DRL converges more quickly and is highly robust. To approximate the state/action/value-function in DRL, a three-layer, ReLU-activated, neural network has been used.

\begin{figure}[t]
    \centering
    \includegraphics[width=0.95\linewidth]{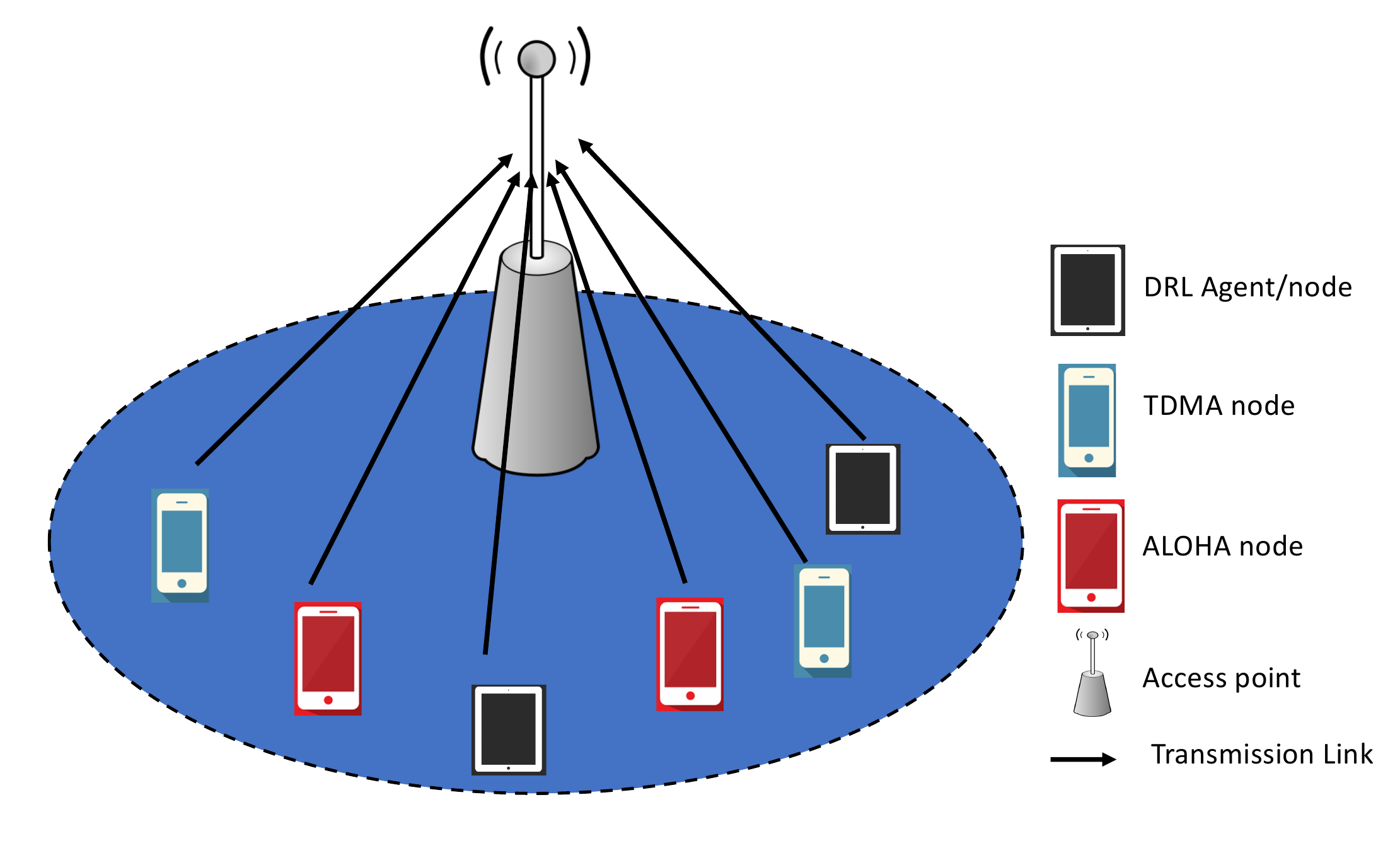}
    \caption{A mix of many nodes and DRL nodes in a multiple access heterogeneous system.}
    \label{fig:DRL_het}
\end{figure}

In \cite{yu2019deep}, the authors proposed a DRL-based MAC protocol known as DLMA, see Fig. \ref{fig:DRL_het}. DLMA considers time slot systems and the problem of dividing time slots between heterogeneous wireless networks. A key characteristic of DLMA is while operating in a heterogeneous environment, it can learn how to reach an overall goal through an array of state/action/reward monitoring. Particularly, without knowing the detailed operating mechanisms of coexisting MACs, DLMA achieves near-optimal performance in relation to the goal. Experiments have been performed to evaluate the DLMA approach against existing RL-based approaches. The results show that DRL converges to the perfect throughput largely greatly faster than RL. Results also show that the DRL node can learn the proper strategy to reach the best throughputs regardless of the MAC protocol type and the number of other nodes.

%\begin{table*}[t]
%  \centering
%  \caption{Taxonomy of MAC protocols in the last decade}
%  \renewcommand{\arraystretch}{1.5}
%  \begin{tabular} {l|c|c|c}
%  \hline \hline
%    Year & Ref. & Taxonomy & Description \\ \hline 
%    2010 & \cite{wang2010cross} & HMAC & Hybrid MAC \\ \hline 
%    2011 & \cite{hu2011load} & LA-MAC & Load Adaptive - MAC \\ \hline 
%    2011 & \cite{ shah2011ddh} & DDHMAC & Dynamic-Decentralized Hybrid MAC \\ \hline 
%    2013 & \cite{ sha2013self } & SAML & Self-Adapting MAC \\ \hline 
%    2015 & \cite{ alvi2015enhanced} & BSMAC & Bitmap-Assisted Shortest Job MAC \\ \hline 
%    2016 & \cite{alvi2016best} & BEST-MAC & Bitmap-Assisted Efficient \& Scalable TDMA based MAC   \\ \hline 
%    2017 & \cite{ngo2017user} & TSCH & Time Slotted Channel Hopping \\ \hline 
%    2019 & \cite{yu2019deep} & DLMA & Deep Learning MAC \\ \hline \hline
%  \end{tabular}
%  \vspace{5pt}
%  \label{tab:Taxonomy}
%\end{table*}

%####################################################
%           Security
%####################################################
\section{Security}
\label{sec:security}

In cybersecurity, Intrusion Detection Systems (IDS) are used for predicting and detecting threats before they lead to larger security incidents. Two main classes of IDS are widely used, namely Signature-based detection and anomaly-based detection. In signature-based detection, a database of various known attack signatures will be maintained, while anomaly-based detection examines and detects intruder behavior by reviewing past activity and checking for deviations from normal traffic behavior. IDS uses various ML approaches specifically, classification techniques to detect cyberattacks.

The authors in \cite{abdulhammed2018enhancing} consider multi-class classification for the Aegean Wi-Fi Intrusion Dataset (AWID) where classes represent $17$ types of the IEEE 802.11 MAC Layer attacks. Four feature groups are considered $32$, $10$, $6$, and $5$ based on various feature selection and reduction algorithms. Firstly, based on reliable and accurate manual feature selection and recommendations from prior works, a total of 32 attributes were selected. The Correlation Feature Selection (CFS) with the Best First Search (BFS) method of forward direction is used to evaluate the feature importance and select the $10$ highest correlated attributes. For the second feature group, Harmony Search (HS) algorithm with the \textit{cost-sensitive subset evaluator} is also employed to evaluate the features and only seven attributes are selected. Finally, the CFS algorithm along with the Harmony Search technique is adopted for the feature selection and dimensionality reduction process, and five attributes are selected. Experiments are conducted to evaluate the effect of the feature selection/reduction on the performance of the classifiers. Seven well-known classification algorithms (i.e., J48, OneR, Naive Bayes, Random Forest, Simple Logistic, Bagging, and Multi-Layer Perceptron) had been evaluated through the selected feature group sets. The results show that, in terms of accuracy and processing time, optimal attribute selection/reduction leads to better results, which is critical for real-time applications.

%\begin{table*}[t]
%  \centering
%  \caption{Taxonomy of MAC protocols in the last decade}
%  \renewcommand{\arraystretch}{1.5}
%  \begin{tabular} {l|c|c|c}
%  \hline \hline
%   Year & Ref. & Taxonomy & Description \\ \hline 
%    2010 & \cite{wang2010cross} & HMAC & Hybrid MAC \\ \hline 
%    2011 & \cite{hu2011load} & LA-MAC & Load Adaptive - MAC \\ \hline 
%    2011 & \cite{ shah2011ddh} & DDHMAC & Dynamic-Decentralized Hybrid MAC \\ \hline 
%    2013 & \cite{ sha2013self } & SAML & Self-Adapting MAC \\ \hline 
%    2015 & \cite{ alvi2015enhanced} & BSMAC & Bitmap-Assisted Shortest Job MAC \\ \hline 
%    2016 & \cite{alvi2016best} & BEST-MAC & Bitmap-Assisted Efficient \& Scalable TDMA based MAC %  \\ \hline 
%    2017 & \cite{ngo2017user} & TSCH & Time Slotted Channel Hopping \\ \hline 
%    2019 & \cite{yu2019deep} & DLMA & Deep Learning MAC \\ \hline \hline
%  \end{tabular}
%  \vspace{5pt}
%  \label{tab:Taxonomy}
%\end{table*}
%#########################################
\section{Conclusion}
Over the past decade, there has been a marked advancement in the field of Machine Learning (ML) and Deep Learning (DL) techniques. These developments have had a significant impact on a wide range of industries. In the realm of wireless communications, the utilization of ML techniques has been implemented to enhance various Medium Access Control (MAC) protocols. This is evidenced by the growing number of publications that utilize the proven capabilities of ML to extend the existing MAC techniques. The present study conducts a survey of ML-based MAC techniques from 2012 to 2022. Additionally, this survey provides an in-depth tutorial on ML techniques, various MAC protocols, and design issues related to MAC protocols. The survey also presents a multi-dimensional taxonomy for the surveyed work, which is based on three main criteria: 1) the machine learning technique adopted in the work, 2) the application area, and 3) the objective of the MAC protocol.

\bibliographystyle{./bibliography/IEEEtran}
\bibliography{./bibliography/IEEEabrv,./bibliography/Ref}

\end{document}